\documentclass[final,seceqn]{elsart1p}

 \usepackage{graphicx}

\usepackage{amsmath,amssymb,amsfonts,bm}

\journal{Nuclear Physics B}
\begin{document}

\begin{frontmatter}

\title{Critical Casimir amplitudes for $n$-component $\phi^4$ models with $O(n)$-symmetry breaking quadratic boundary terms }

\author{H.~W. Diehl and Daniel Gr\"uneberg}

\address{ Fachbereich Physik, Universit{\"a}t Duisburg-Essen, 47048
  Duisburg, Germany}

\begin{abstract}
 Euclidean $n$-component $\phi^4$ theories whose Hamiltonians are $O(n)$ symmetric except for  quadratic symmetry breaking boundary terms are studied in the film geometry $\mathbb{R}^{d-1}\times[0,L]$. The boundary terms imply the Robin boundary conditions $\partial_n\phi_\alpha=\mathring{c}^{(j)}_\alpha\phi_\alpha$ at the boundary planes $\mathfrak{B}_{j=1}$  at $z=0$  and $\mathfrak{B}_{j=2}$ at $z=L$. Particular attention is paid to the cases in which  $m_j$ of the $n$ variables $\mathring{c}^{(j)}_\alpha$ associated with plane $\mathfrak{B}_j$ take the special value $\mathring{c}_{m_j\text{-sp}}$ corresponding to critical enhancement while the remaining ones are larger and hence subcritically enhanced. Under these conditions, the semi-infinite system with boundary plane $\mathfrak{B}_j$ has a multicritical surface-bulk point, called $m_j$-special, at which an $O(m_j)$ symmetric critical surface phase coexists with the $O(n)$ symmetric bulk phase, provided $d$ is sufficiently large. The  $L$-dependent part of the reduced free energy per cross-section area behaves asymptotically as $\Delta_C/L^{d-1}$ as $L\to\infty$ at the bulk critical point. The Casimir amplitudes $\Delta_C$ are determined for small $\epsilon=4-d$ in the general case where $m_{c,c}$ components $\phi_\alpha$ are critically enhanced at both boundary planes, $m_{c,\mathrm{D}}+m_{\mathrm{D},c}$ components are enhanced at one plane but satisfy asymptotic Dirichlet boundary conditions at the respective other, and  the remaining $m_{\mathrm{D},\mathrm{D}}$ components  satisfy asymptotic  Dirichlet boundary conditions at both $\mathfrak{B}_j$.  Whenever $m_{c,c}>0$, the corresponding small-$\epsilon$ expansions involve, besides integer powers of $\epsilon$, also fractional powers $\epsilon^{k/2}$ with $k\ge 3$ modulo powers of logarithms. Results  to order $\epsilon^{3/2}$ are given for general values of $m_{c,c}$, $m_{c,\mathrm{D}}+m_{\mathrm{D},c}$, and $m_{\mathrm{D},\mathrm{D}}$. These are used  to estimate the Casimir amplitudes $\Delta_C$ of the three-dimensional Heisenberg systems  with surface spin anisotropies for the cases with $(m_{c,c},m_{c,\mathrm{D}}+m_{\mathrm{D},c})=(1,0)$, $(0,1)$, and $(1,1)$.
  \end{abstract}

\end{frontmatter}

\section{Introduction} \label{sec:intro}
The theory of finite size effects at continuous phase transitions \cite{Fis71}  has advanced considerably during the past decades \cite{Bar83,Bin87,Pri90,BDT00,Kre94}.\footnote{Extensive lists of references can be found in the review articles \cite{Bar83,Bin87,Pri90,BDT00,Kre94}.}  There is growing evidence that the  finite-size scaling functions it involves become independent of microscopic details provided appropriate variables (scaling fields) are chosen and certain nonuniversal metrical coefficients are fixed \cite{Doh08,DC09}.\footnote{For recent discussions of these issues, see references \cite{Doh08} and \cite{DC09}.}  Universal finite-size properties generally depend on (i) all gross bulk properties that determine the bulk universality class (such as space dimension $d$,  type of order parameter, and gross features of the interactions), (ii) gross features of the boundary (such as large-scale boundary conditions and other properties that determine their boundary critical behavior), and (iii)  geometric properties such as shape or curvature. 
 
 Here we shall be concerned with $d$-dimensional  systems in a $\infty^{d-1}\times L$ film geometry. We have in mind systems of classical $n$-vector spins whose bulk critical behavior is described by a standard $O(n)$ invariant  $\phi^4$ model. Generalizations of this model to films have been studied in a number of papers \cite{KD91,KD92a,KD92b,SD08}. A finite-size quantity of primary interest is the free energy density. The total free energy per cross-section area $A$ and thermal energy $k_BT$ of such systems  can be decomposed as
 \begin{equation}
 \label{eq:Fsplit}
f_L(T)\equiv\lim_{A\to\infty}\frac{F}{Ak_BT}=L\,f_b(T)+f_s(T)+f_{\mathrm{res}}(L;T)
\end{equation}
 in the limit $A\to\infty$, where $f_b$ and $f_s$ are the reduced bulk free-energy density and the reduced surface excess free-energy density, respectively, both of which are independent of $L$.  The  remaining $L$-dependent contribution is the residual free-energy density $f_{\mathrm{res}}(L;T)$. We have kept only the thermodynamic field $T$, assuming the absence of magnetic fields. 
 
 For films whose interactions are homogeneous throughout the sample except for local modifications near the top and bottom planes of the film, $f_b$ depends merely on the bulk interaction constants. By contrast, $f_s$ and $f_{\mathrm{res}}$ additionally depend on the modified (``surface'') interaction constants at both boundary planes. 
According to finite-size scaling theory, the singular part of $f_{\mathrm{res}}(L;T)$ should have a scaling form which directly at the bulk critical point $T=T_{c,\infty}$, yields the asymptotic large-$L$  behavior
 \begin{equation}
\label{eq:frescrit}
f^{\mathrm{sing}}_{\mathrm{res}}(L;T_{c,\infty})\approx \Delta_C^{(\wp)}\,L^{-(d-1)}\;.
\end{equation}
The associated  amplitudes $\Delta_C^{(\wp)}$, called critical Casimir amplitudes, depend on the boundary conditions $\wp$ that hold asymptotically on large length scales. They have been computed by means of the mean-field approximation, approximate re\-norm\-al\-iza\-tion-group (RG) methods, exact model solutions, computer simulations, and conformal field theory for periodic ($\wp=\mathrm{P}$) and  antiperiodic ($\wp=\mathrm{AP}$) boundary conditions, as well as a variety of large-scale boundary conditions one encounters for lattice systems with free boundary conditions \cite{KD91,KD92a,NI85,INW86,KL96,Kre97,DK04,DDG06,Huc07,VGMD07}. 

The latter large-scale boundary conditions may be characterized via the universality classes pertaining to the boundary critical behavior at the bulk critical point of semi-infinite systems bounded by a pair of surface planes of the corresponding kinds at $z=0$ and $z=\infty$. If the Hamiltonian is $O(n)$ symmetric, three distinct familiar types of such ``surface universality classes'' may be distinguished: those associated with the ordinary, special, and extraordinary transitions \cite{Bin83,Die86a,Die97}. Which one of these transitions occurs, depends on whether the enhancement of the pair interactions at the surface relative to those inside the film is weaker than, equal to, or stronger than a threshold (the ``critical enhancement''). For those values of $n$ and $d$ for which all three kinds of surface transitions are possible in semi-infinite geometry, we can specify the boundary conditions $\wp$ by a pair $(a,b)$ of surface universality classes $a,b=\text{ord}$, $\text{sp}$, and $\text{ex}$. Since for supercritical enhancement, a transition to a surface ordered, bulk disordered  phase occurs at a temperature $T_{s,\infty}>T_{c,\infty}$, the symmetry is spontaneously broken at the extraordinary transition. Therefore,  boundary conditions $(a,b)$ with $a=\text{ex}$ or $b=\text{ex}$ differ qualitatively from those involving only $a,b=\text{ord}$ and $\text{sp}$ inasmuch as the former break the internal symmetry $O(n)$ down to $O(n-1)$, whereas the latter preserve it. Symmetry breaking large-scale boundary conditions are also encountered for arbitrary (sub- or supercritical) enhancement when boundary contributions linear in $\bm{\phi}$ are present in the Hamiltonian. Such a situation is generically encountered in the scalar case of one-component and binary mixed fluids bounded by walls, i.e., the so-called normal surface transition \cite{Die86a,Die97,BD94,Die94a}. The explicit symmetry breaking (by linear boundary terms in the Hamiltonian) one has there is physically distinct from the spontaneous symmetry breaking that occurs in the surface ordered, bulk disordered phase and hence is present at the extraordinary transition. Nevertheless, the asymptotic boundary conditions are the same in both cases; they correspond to fixed nonzero values of the order parameter at the surface.

In this paper we will be concerned with the Casimir effect under symmetry breaking boundary conditions of a different kind. Our aim is to generalize previous work on the Casimir effect in models with $O(n)$ symmetric Hamiltonian \cite{KD92a} by including
\emph{quadratic} symmetry-breaking boundary terms. Our motivation is the following. The spontaneous breaking of the continuous $O(n)$ symmetry that the extraordinary transition involves requires the surface dimension $d-1$ to be larger than $2$. This means that at $d\le 3$ neither the extraordinary nor the special surface transitions are possible for $O(n)$ symmetric Hamiltonians and free boundary conditions.%
\footnote{The case $n=2$ is special since a surface phase with quasi-long-range order is possible at $d=3$. Thus a surface-bulk multicritical point  at which a line of surface transitions of Kosterlitz-Thouless type reaches the bulk critical temperature may be anticipated. We will not embark on an investigation of this possibility.}
By contrast, normal transitions are possible whenever $d$ exceeds the lower critical dimension $d_*=2$ of the $O(n)$ bulk model. Thus, for three-dimensional isotropic $n$-vector models the possibility of a special transition at finite critical enhancement  does not exist.%
\footnote{In a three-dimensional semi-infinite lattice model whose spins are coupled by ferromagnetic nearest-neighbor bonds $K_{\bm{x},\bm{x}'}$ of strengths $K_1$ and $K$ depending on whether both sites $\bm{x}$ and $\bm{x}'$ belong to the surface layer or not, the entire line of extraordinary transitions including the surface-bulk multicritical point associated with the special transition should degenerate into a point with $(K_1/K)_{\text{sp}}=\infty$.}  However, in systems whose interactions are known to be isotropic in the bulk (such as the isotropic Heisenberg model) it is quite common that surface spin anisotropies occur, which entail quadratic symmetry-breaking boundary terms in the Hamiltonian. When they are allowed, the picture changes and becomes much richer. Suppose the boundary terms at plane $\mathfrak{B}_j$ are isotropic in an $m_j$-dimensional subspace with $1\le m_j<n$. Then a surface transition breaking this $O(m_j)$ symmetry can occur if the surface coupling associated with this easy manifold is sufficiently strong and $d-1$ exceeds the respective lower critical (bulk) dimensions $2$ and $1$ for  continuous $(m_j>1)$ and $\mathbb{Z}_2$ symmetry, respectively. Consequently, anisotropic analogs of the special transition, so-called $m_j$-special transitions \cite{DE82,DE84},  occur  in  the corresponding semi-infinite systems when the enhancement of this coupling takes a special value while the remaining surface couplings are weaker.  For $d=3$, all cases of critically enhanced easy-axis spin anisotropies --- namely, those where either $m_{1}$ or $m_{2}$ or else both are unity --- are of interest. 

One could also consider the cases where some or all  remaining surface couplings are stronger (i.e., supercritically enhanced). Following Ref.~\cite{DE84}, we shall exclude this possibility for the sake of simplicity in all our explicit calculations below. However, we shall investigate the general case of a $(d=4-\epsilon)$-dimensional film whose  surface interactions are such that the respective semi-infinite  systems with surface planes $\mathfrak{B}_1$ and $\mathfrak{B}_2$ would undergo  $m_{1}$- and $m_2$-special surface transitions of this kind with $0\le m_j\le n$ at $T_{c,\infty}$. The weak (not critically enhanced) surface interaction constants are driven to a fixed point at which those  components of $\bm{\phi}$ with which they are associated  satisfy Dirichlet boundary conditions. Extending our previous analyses \cite{DGS06,GD08}, we shall determine the series expansions of  the corresponding Casimir amplitudes $\Delta_C^{(\wp)}$  in $\epsilon=4-d$ to order $\epsilon^{3/2}$. Whenever components $\phi_\alpha$ exists whose surface interaction constants are critically enhanced on both boundary planes, these series expansions involve fractional powers $\epsilon^{k/2}$ with $k\ge 3$ (modulo powers of logarithms) besides integer powers of $\epsilon$.

The remainder of this paper is organized as follows. In Section~\ref{sec:model} the model is introduced. Section~\ref{sec:bg} sets the stage for the subsequent RG analysis, providing some necessary  background on the model's renormalization, on anisotropic special transitions in semi-infinite systems, their fixed points, and the modified RG-improved perturbation theory required in those cases where zero modes are present at $T_{c,\infty}$ for finite $L$ in Landau theory. Section~\ref{sec:CA} describes our calculation of critical Casimir amplitudes and presents  our results. Section~\ref{sec:concl} contains a discussion of our main findings and concluding remarks. Finally, there are two appendices with technical details. 
 
\section{Model } 
  \label{sec:model}

We consider a $\phi^{4}$ model for an $n$-component order-parameter field $\bm{\phi}(\bm{x})=(\phi_\alpha(\bm{x});\alpha=1,\ldots,n)$ defined on the  $d$-dimensional slab $\mathfrak{V}=\mathbb{R}^{d-1}\times[0,L]$. Introducing Cartesian coordinates $x_j$, we write the position vector $\bm{x}=(x_1,\ldots,x_d)$ as $\bm{x}=(\bm{y},z)$, where $\bm{y}=(x_1,\ldots,x_{d-1})$ is the $(d-1)$-dimensional lateral component, while $z\equiv x_d$ is the coordinate across the slab. It is understood that periodic boundary conditions are chosen along all $y$ directions, so that the boundary $\partial\mathfrak{V}$ consists of the union  $\mathfrak{B}_1\cup\mathfrak{B}_2$ of the two hyperplanes $z=0$  and $z=L$.
 
We assume that long-range interactions are either absent or can be ignored, and that the modifications of the interactions in the vicinity of the boundary planes $\mathfrak{B}_1$ and $\mathfrak{B}_2$ are short ranged. Under these assumptions a mesoscopic description in terms of a local field theory with a Hamiltonian of the form
 \begin{equation} \label{eq:Hform}
\mathcal{H}[\bm{\phi}]=\int_{\mathfrak{V}}\mathcal{L}_{\mathfrak{V}}(\bm{x})\,\mathrm{d} V+\int_{\partial\mathfrak{V}}\mathcal{L}_{\partial\mathfrak{V}}(\bm{x})\,\mathrm{d} A
\end{equation}
is possible \cite{Die86a,Die97}. Here the volume and surface elements are given by $\mathrm{d} V=\mathrm{d}^dx$ and $\mathrm{d} A=\mathrm{d}^{d-1}y$, respectively. We  orient the boundary $\partial\mathfrak{V}$ such that the normal $\bm{n}$ points into the interior of $\mathfrak{V}$. 

The bulk and surface densities $\mathcal{L}_{\mathfrak{V}}$ and $\mathcal{L}_{\partial\mathfrak{V}}$ depend on $\bm{\phi}$ and its spatial derivatives. We choose for the former the $O(n)$ invariant expression
\begin{equation} \label{eq:Lb}
\mathcal{L}_{\mathfrak{V}}[\bm{\phi}]=\frac{1}{2}\,\sum_{\alpha=1}^{n}(\nabla\phi_{\alpha})^{2}+\frac{\mathring{\tau}}{2}\,\phi^{2}+\frac{\mathring{u}}{4!}\,\phi^{4}\,,
\end{equation}
where $\phi= |\bm{\phi} |$ is the absolute value. The surface density is taken to consist of general terms quadratic in $\bm{\phi}$ that break the $O(n)$ symmetry of $\mathcal{L}_{\mathfrak{V}}$; we write it as
\begin{equation}
 \label{eq:L1}
\mathcal{L}_{\partial\mathfrak{V}}[\bm{\phi}]=\frac{1}{2}\sum_{\alpha=1}^{n}\mathring{c}_{\alpha}(\bm{x})\phi_{\alpha}(\bm{x})^{2}\;,
\end{equation}
where the surface enhancement variables $c_{\alpha}(\bm{x})$ are allowed to
take different, yet position-independent, values on the planes $\mathfrak{B}_1$ and $\mathfrak{B}_2$. In other words, we assume that
\begin{equation}
 \label{eq:c12}
\mathring{c}_{\alpha}(\bm{x})=\mathring{c}^{(j)}_{\alpha}\quad \mbox{ for }\bm{x}\in\mathfrak{B}_j\;.
\end{equation}

From the boundary terms of the classical equations of motion (implied by $\delta\mathcal{H}=0$), one obtains the boundary conditions
\begin{equation}
 \label{eq:bc}
\partial_n\phi_\alpha(\bm{X})=\mathring{c}^{(j)}_\alpha\phi_\alpha(\bm{X})\quad \text{for }\bm{X}\in\mathfrak{B}_j\;.
\end{equation}
These mesoscopic (Robin) boundary conditions hold beyond Landau theory in an operator sense --- i.e., inside of averages  --- for the bare regularized theory \cite{Die86a,Die97}. In our actual calculation described in Section~\ref{sec:casicalc} we shall use dimensional regularization, although we shall occasionally comment on how the required counterterms depend on a large-momentum cutoff if such a regularization scheme were used instead.

From the boundary conditions~\eqref{eq:bc} and the surface density~\eqref{eq:L1} it is clear that  $\phi_\alpha$   will get strongly suppressed at $\mathfrak{B}_j$ when  $\mathring{c}^{(j)}_\alpha$ takes large positive values. In the limit $\mathring{c}^{(j)}_\alpha\to \infty$, Eq.~\eqref{eq:bc}  turns into a Dirichlet boundary condition for $\phi_\alpha$ on $\mathfrak{B}_j$. These observations tie in with the known fact that the deviation of $-\mathring{c}^{(j)}_\alpha$ from a reference value is a measure of how much the pair interaction on $\mathfrak{B}_j$ is enhanced relative to its bulk counterpart \cite{Bin83,Die86a}. For our subsequent analysis it is also important to remember that the mesoscopic boundary condition~\eqref{eq:bc} must be carefully distinguished from the large-scale boundary conditions mentioned in the Introduction. Just as other coupling constants of the mesoscopic theory, the variables $\mathring{c}^{(j)}_\alpha$ depend on the length scale to which one has coarse-grained. Integrating out the degrees of freedom up to a larger length scale generally produces modified values of the interaction constants. In a field-theoretic RG approach, their renormalized analogs become running variables. The asymptotic large-scale boundary conditions are associated with RG fixed points \cite{Die86a,Die97}. To work out the details for our generalized  model with surface spin anisotropies, we first must recall some background on semi-infinite systems involving quadratic symmetry breaking boundary terms. 

\section{Background}
 \label{sec:bg}
 
\subsection{Free propagators and boundary conditions}
Consider the ($N+M$)-point cumulants involving $N$ interior points $\bm{x}_{j}\notin\partial\mathfrak{V}$
and $M$ boundary points $\bm{X}_{k}\in\partial\mathfrak{V}$ in $d=4-\epsilon$ dimensions,
\begin{equation}
   \label{eq:GNMdef}
G_{\alpha_{1},\ldots,\beta_{M}}^{(N,M)}(\bm{x}_{1},\ldots,\bm{X}_{M})
=\bigg\langle\prod_{j=1}^{N}\phi_{\alpha_{j}}(\bm{x}_{j})
\prod_{k=1}^{M}\phi_{\beta_{k}}(\bm{X}_{k})\bigg\rangle^{\mathrm{cum}}\;.
\end{equation}
In the disordered phase, the Fourier $\bm{y}$-transform 
\begin{equation}
\label{eq:parFT}
G_{L,\alpha}(\bm{p};z_1,z_2)=\int \mathrm{d}^{d-1}y\,G_{L,\alpha}(\bm{y}_1,z_1;\bm{y}_2,z_2)\,\mathrm{e}^{-i\bm{p}\cdot\bm{y}_{12}}
\end{equation}
of the $\alpha=\beta$ element of the free propagator $G_{L,\alpha}(\bm{x}_1;\bm{x}_2)\,\delta_{\alpha\beta}$ satisfies the differential equation
\begin{equation}
\label{eq:Gfreeeq}
(-\partial_z^2+p^2+\mathring{\tau})G_{L,\alpha}(\bm{p};z_1,z_2)=\delta(z_1-z_2),
\end{equation}
subject to the boundary conditions
\begin{eqnarray}
 \label{eq:Gbc}
\left.  \partial_zG_{L,\alpha}(\bm{p};z,z_2)\right|_{z=0}=\mathring{c}^{(1)}_\alpha G_{L,\alpha}(\bm{p};0,z_2),\nonumber\\
 \left.  -\partial_zG_{L,\alpha}(\bm{p};z,z_2)\right|_{z=L}=\mathring{c}^{(2)}_\alpha G_{L,\alpha}(\bm{p};L,z_2),
\end{eqnarray}
where $\bm{x}_j=(\bm{y}_j,z_j)$ and $\bm{y}_{12}=\bm{y}_1-\bm{y}_2$. The solution can be expressed in terms of the eigenvalues $[\hat{k}_r(c_1,c_2)]^2$ and  eigenfunctions $ \varphi_r(\zeta|c_1,c_2)$ of the operator $-\partial_\zeta^2$ on the interval $[0,1]$ obeying  the boundary conditions
\begin{eqnarray}
 \label{eq:varphibc}
 \varphi_r'(0|c_1,c_2)&=&c_1\,\varphi_r(0|c_1,c_2)\,,\nonumber\\  - \varphi_r'(1|c_1,c_2)&=&c_2\,\varphi_r(1|c_1,c_2),
\end{eqnarray}
where $c_1,c_2\in[0,\infty)$ and $\zeta=z/L$ \cite{RS02,Sch08}. This gives
\begin{eqnarray}
\label{eq:Gfreespecrep}
G_{L,\alpha}(\bm{p};z_1,z_2)&\equiv&  G_{L,\alpha}(\bm{p};z_1,z_2|\mathring{\tau},c^{(1)}_\alpha,c^{(2)}_\alpha)\nonumber\\ &=&\frac{1}{L}\sum_{r}\frac{\varphi_r\big(z_1/L|L\mathring{c}^{(1)}_\alpha,L\mathring{c}^{(2)}_\alpha\big)\,\varphi^*_r\big(z_2/L|L\mathring{c}^{(1)}_\alpha,L\mathring{c}^{(2)}_\alpha\big)}{p^2+\mathring{\tau}+\big[\hat{k}_r\big(L\mathring{c}^{(1)}_\alpha,L\mathring{c}^{(2)}_\alpha\big)/L\big]^2}\;.
\end{eqnarray}

The eigenfunctions $\varphi_r(\zeta=z/L|c_1,c_2)$ are phase shifted cosine functions $\propto\cos(\hat{k}_r\zeta+\vartheta_r)$ whose phase shift $\vartheta_r\equiv\vartheta_r(c_1,c_2)$ follows from the boundary conditions at $z=0$. The boundary condition at $z=L$ yields an equation for the eigenvalues $k_r^2=(\hat{k}_r/L)^2$, which for general values of $c_1$ and $c_2$ is transcendental. Details can be found in Refs.~\cite{RS02} and \cite{Sch08}, but will not be repeated here since we shall need the explicit expressions for the free propagators $G_{L,\alpha}(\bm{p};z_1,z_2)$ and the eigenfunctions only for the  special choices $\mathring{c}^{(j)}_\alpha=0,\infty$.  For those, the free propagators satisfy Dirichlet (D) or Neumann (N) boundary conditions on $\mathfrak{B}_j$, depending on whether $\mathring{c}^{(j)}_\alpha=\infty$ or $\mathring{c}^{(j)}_\alpha=0$. The eigenfunctions and eigenvalues then simply become
\begin{align} \label{eq:varphiAB}
\varphi^{(\mathrm{N,N)}}_r(\zeta)&\equiv\varphi(\zeta|0,0)=\sqrt{2-\delta_{r,0}}\cos(\hat{k}_r\zeta),&\hat{k}_r&=r\pi,&r&=0,1,\dotsc,\infty,\nonumber \\[\medskipamount]
\varphi^{(\mathrm{D,D)}}_r(\zeta)&\equiv\varphi(\zeta|\infty,\infty)=\sqrt{2}\sin(\hat{k}_r\zeta),&\hat{k}_r&=r\pi,&r&\in \mathbb{Z}, \nonumber \\[\medskipamount]
\varphi^{(\mathrm{N,D)}}_r(\zeta)&\equiv\varphi(\zeta|0,\infty)=\sqrt{2}\cos(\hat{k}_r\zeta),&\hat{k}_r&=\frac{2r+1}{2}\pi,&r&=1,2,\dotsc,\infty,\nonumber \\[\medskipamount]
\varphi^{(\mathrm{D,N)}}_r(\zeta)&\equiv\varphi(\zeta|\infty,0)=\sqrt{2}\sin(\hat{k}_r\zeta),&\hat{k}_r&=\frac{2r+1}{2}\pi,&r&=1,2,\dotsc\infty.
\end{align}

The respective free propagators can be written in terms of $G_\infty^{(d)}(\bm{p};z)$, the Fourier $\bm{y}$-transform  of the free bulk propagator
\begin{equation}
  \label{eq:bulkpropag}
G_{\infty}^{(d)}(\bm{x}|\mathring{\tau}) =  \int_{\bm{q}}^{(d)} \frac{\mathrm{e}^{i\bm{q}\cdot\bm{x}}}{q^{2}+\mathring{\tau}}
 =(2\pi)^{-d/2} \left(\mathring{\tau}/x^{2}\right)^{(d-2)/4}\, K_{(d-2)/2}\big(x\sqrt{\mathring{\tau}}\big)
 \end{equation}
in $d$ dimensions, where $\int_{\bm{q}}^{(d)}\equiv(2\pi)^{-d}\int_{\mathbb{R}^d}\mathrm{d}^dq$ is a convenient short-hand. One finds
\begin{eqnarray}
 \label{eq:GAB}
G^{\text{(A,B)}}_L(\bm{p};z_1,z_2)&=&\sum_{j=1}^\infty(-1)^{j(1-\delta_{\text{AB}})}\big[G^{(d)}_\infty(\bm{p};z_1-z_2+2jL) \nonumber\\ &&\strut -(-1)^{\delta_{\text{AN}}\delta_{\text{BN}}}\,G^{(d)}_\infty(\bm{p};z_1+z_2+2jL)\big],\;\;\text{A,B}=\text{N,D}.\nonumber\\
	\end{eqnarray}
These results can be obtained directly by means of the method of images. Alternatively, one can start from Eq.~\eqref{eq:Gfreespecrep}, substitute the eigenfunctions~\eqref{eq:varphiAB}, and then use  Poisson's summation formula (see, e.g., Eq.~(4.8.28) of Ref.~\cite{MF53}) 
\begin{equation}
 \label{eq:PoissonSumFormula}
\frac{1}{L}\sum_{r=-\infty}^{\infty}f(r\pi/L)=
\sum_{j=-\infty}^{\infty}\int_{-\infty}^{\infty}\frac{\mathrm{d}{k}}{\pi}f(k)\mathrm{e}^{
  \mathrm{i} kj 2L}\;.
\end{equation}
Closed-form expressions for $G^{\text{(A,B)}}_L(\bm{p};z_1,z_2)$ will not be needed but may be gleaned from Eq.~(B5) of Ref.~\cite{DC09}; its right-hand side, with the frequency variable $\omega$ appearing  there set to zero,  gives $ G_{L,\alpha}(\bm{p};z_1,z_2|\mathring{\tau},c^{(1)}_\alpha,c^{(2)}_\alpha)$ for general non-negative values of $c^{(1)}_\alpha$ and $c^{(2)}_\alpha$. 

\subsection{Renormalization by expansion about the isotropic special point}
\label{sec:isoren}

Upon expanding in the variables $\mathring{c}^{(j)}_\alpha$, we can work with the free Neumann-Neumann propagator $G^{(\text{N,N})}_L(\bm{x}_1; \bm{x}_2)$. This has integrable ultraviolet (UV) singularities at coinciding points originating from its bulk contribution $G_\infty^{(d)}(\bm{x}_1-\bm{x}_2)$ and additional ones localized on the boundary planes $\mathfrak{B}_1$ and $\mathfrak{B}_2$ due to two other summands in Eq.~\eqref{eq:GAB}. As is well known, these UV singularities induce divergences in the Feynman integrals for $G_{\alpha_1,\dotsc,\beta_M}^{(N,M)}$ which can be absorbed  in a systematic fashion by appropriate reparametrizations when $d\le 4$. In addition to bulk counterterms, counterterms with support on the boundary planes are needed \cite{Die86a,Die97,DD80,Sym81,DD81a,DD81b,DD83a}. They can be chosen to correspond to the reparametrizations known from the bulk and semi-infinite analogs of the model. We use the bulk reparametrizations
\begin{equation}
\label{eq:brep}
\bm{\phi}  = Z_{\phi}^{1/2}\,\bm{\phi}_{R},\quad
\delta\mathring{\tau} \equiv \mathring{\tau}-\mathring{\tau}_{c,\infty} =  Z_{\tau}\,\mu^{2}\tau, \quad
\mathring{u}\, N_{d} = \mu^{\epsilon}\, Z_{u}\, u,
\end{equation}
to introduce a renormalized field $\bm{\phi}_R$, renormalized variable $\tau$, and coupling constant $u$, where $\mu$ is an arbitrary momentum scale. Following Ref.~\cite{GD08} and \cite{DC09}, we  
choose the factor that is absorbed in the renormalized coupling constant as 
\begin{equation}
N_{d}  =  \frac{2\,\Gamma(3-d/2)}{(d-2)(4\pi)^{d/2}}
  =  \frac{1}{16\pi^{2}}\bigg[1+\frac{1-C_{E}+\ln(4\pi)}{2}\,\epsilon+O(\epsilon^{2})\bigg].
 \end{equation}

Owing to the presence of the $O(n)$-symmetry breaking boundary terms, the usual surface reparametrizations \cite{DD83a,Die86a} must be generalized as expounded in Refs.~\cite{DE84} and \cite{DE85}.  Accordingly, we introduce renormalized surface enhancement variables $c_\alpha^{(j)}$ and renormalized boundary fields $\bm{\phi}|_{\partial\mathfrak{B}}^{R}$ via
 \begin{equation}
 \label{eq:srep}
\delta\mathring{c}_\alpha^{(j)}   =  \mu\sum_{\beta=1}^n Z_{\alpha\beta}\, c_\beta^{(j)},\quad \delta\mathring{c}_\alpha^{(j)}\equiv \mathring{c}_\alpha^{(j)}-\mathring{c}_{\text{sp}}\,,\quad
{\bm{\phi}}|_{\partial\mathfrak{B}} =  (Z_{\phi}Z_{1})^{1/2}\,\bm{\phi}|_{\partial\mathfrak{B}}^{R}\,,
\end{equation}
where $\bm{\phi}|_{\partial\mathfrak{B}}$ means $\bm{\phi}$, taken
at a boundary point $\bm{X}$, and $Z_{\alpha\beta}$ has the form
\begin{equation}
Z_{\alpha\beta}=[Z_c+(n\delta_{\alpha\beta}-1)Z_f]/n.
\end{equation}
We fix the renormalization factors $Z_\phi$, $Z_\tau$, $Z_u$, $Z_1$, $Z_c$, and $Z_f$ by minimal subtraction of poles in $\epsilon=4-d$. Then they agree with the results to order $u^2$ given in Eqs.~(3.142a)--(3.142c), (3.66a), (3.166b) of Ref.~\cite{Die86a} and Eq.~(49) of Ref.~\cite{DE84}.

It is convenient to transform to enhancement variables that transform orthogonally under the RG. Let $\bm{U}=(U_{\alpha\beta})$ be an orthogonal matrix with $U_{n\beta}=n^{-1/2}$. Upon introducing the symmetric deviations $\delta\mathring{c}^{(j)}_\alpha$ and anisotropy variables $\mathring{f}^{(j)}_\alpha$ via
\begin{equation}
 \label{eq:transf}
\sum_{\beta=1}^nU_{\alpha\beta}\,\delta \mathring{c}^{(j)}_\beta= \begin{cases}\mathring{f}^{(j)}_\alpha&\text{ for } \alpha=1,\dotsc,n-1,\\[\medskipamount]
{\displaystyle \delta\mathring{c}^{(j)}=\sum_{\beta=1}^n\delta\mathring{c}^{(j)}_\beta/n}&\text{ for }\alpha=n,
\end{cases}
\end{equation}
as well as anisotropy operators 
\begin{equation}
 \label{eq:Aop}
A_\alpha=\sum_{\beta=1}^{n}U_{\alpha\beta}\phi_\beta^2\,,\quad\alpha=1,2,\dotsc,n-1,
\end{equation}
(which we do not further specify; for explicit expressions, see Ref.~\cite{DE84}), the boundary operators describing the deviations from the isotropic special point become
\begin{equation}
 \label{eq:rewrbt}
\frac{1}{2}\sum_{\alpha=1}^n\delta\mathring{c}^{(j)}_\alpha\,\phi_\alpha^2=\frac{1}{2}\,\delta\mathring{c}^{(j)}\phi^2+\frac{1}{2}\sum_{\beta=1}^{n-1}\mathring{f}^{(j)}_\alpha A_\alpha\,.
\end{equation}
Further, the corresponding linear combinations of the renormalized enhancement variables $c^{(j)}_\alpha$ are related to these bare variables in the simple multiplicative fashion
\begin{equation}
 \label{eq:cfren}
\delta\mathring{c}^{(j)}=\mu Z_c c^{(j)}\;,\qquad\mathring{f}^{(j)}_\alpha=\mu Z_ff^{(j)}_\alpha\,,\;\;\alpha=1,\dotsc,n-1.
\end{equation}

It follows that the renormalized functions $G^{(N,M)}_R=Z_\phi^{-(N+M)/2}Z_1^{-M/2}G^{(N,M)}$ (whose component indices $\alpha_1,\dotsc,\beta_M$ we suppress) satisfy the RG equations
\begin{equation}
 \label{eq:GRGiso}
\left[ \mathcal{D}_\mu+\frac{M+N}{2}\,\eta_\phi+\frac{M}{2}\,\eta_1\right]G^{(N,M)}_R=0
 \end{equation}
 with
 \begin{equation}
\mathcal{D}_\mu=\mu\partial_\mu+\beta_u\partial_u-(2+\eta_\tau)\partial_\tau-(1+\eta_c)\sum_{j=1}^2c^{(j)}\partial_{c^{(j)}}
-(1+\eta_f)\sum_{j=1}^2\sum_{\alpha=1}^{n-1}f^{(j)}\partial_{f^{(j)}},
\end{equation}
where the beta function $\beta_u$ and the exponent functions $\eta_w$ are defined through the $\mu$-derivatives $\partial_{\mu}|_0$ at fixed bare interaction constants,
\begin{equation}
 \label{eq:betaexpfctns}
\beta_u\equiv\mu\partial_{\mu}|_0u=-[\epsilon+\eta_u(u)]u,\qquad\eta_w(u)\equiv\mu\partial_{\mu}|_0\ln Z_w,\;\;w=\phi,\tau,u,1,c,f.
\end{equation}

The characteristics of these RG equations define us running interaction variables $\bar{u}(\ell)$, $\bar{\tau}(\ell)$, $\bar{c}(\ell)$, and $\bar{f}^{(j)}_\alpha(\ell)$ describing their flow under changes $\mu\to\mu \ell$. The flow equations of  $\bar{u}(\ell)$, $\bar{\tau}(\ell)$, and $\bar{c}(\ell)$ and the respective initial conditions can be found in Ref.~\cite[Eq.~(3.79a--c)]{Die86a}; those of the $\bar{f}^{(j)}_\alpha(\ell)$ read
\begin{equation}
\label{eq:flowfs}
\ell \frac{\mathrm{d}}{\mathrm{d} \ell}\bar{f}^{(j)}_\alpha(\ell)=-[1+\eta_f(\bar{u})]\bar{f}^{(j)}_\alpha(\ell),\quad\bar{f}^{(j)}_\alpha(1)=f^{(j)}_\alpha.
\end{equation}
The asymptotic behaviors of the isotropic enhancement $\bar{c}(\ell)$ and the anisotropies $\bar{f}^{(j)}_\alpha(\ell)$ in the large-length-scale limit $\ell\to 0$ are governed by the familiar (surface-bulk) crossover exponent
\begin{equation}
\Phi=\nu[1+\eta_c(u^*)] 
\end{equation}
and the anisotropy crossover exponent (cf.\ Ref.~\cite{DE84})
\begin{equation}
\Psi=\nu[1+\eta_f(u^*)],
\end{equation}
where $u^*=O(\epsilon)$ is the nontrivial root of $\beta_u$ (infrared-stable fixed point). One has
$\bar{c}(\ell)\sim\ell^{-\Phi/\nu}c$ and $\bar{f}^{(j)}_\alpha(\ell)\sim \ell^{-\Psi/\nu}f^{(j)}_\alpha$.

As is well known, renormalization of the bulk and surface free energy densities $f_b$ and $f_s$ requires, in addition to the counterterms  implied by the bulk and surface reparametrizations~\eqref{eq:brep} and \eqref{eq:srep}, also additive counterterms. Owing to the latter, the corresponding RG equations are inhomogeneous. However, the
additive counterterms can be chosen in a  way that their contributions drop out in the residual free energy $f_{\text{res}}$ \cite{Die86a}, which therefore satisfies the homogeneous RG equation
\begin{equation}
 \label{eq:RGEfres}
\mathcal{D}_\mu f_{\text{res}}=0.
\end{equation}
Solving this at the bulk critical point $T=T_{c,\infty}$ (and neglecting corrections to scaling), we arrive at the scaling form
\begin{equation}
f_{\text{res}}(L;T_{c,\infty})\approx D\big(c^{(1)}L^{\Phi/\nu},c^{(2)}L^{\Phi/\nu};\{f^{(j)}_\alpha L^{\Psi/\nu}\} \big)\,L^{-(d-1)}.
\end{equation}
The restriction of the function $D\big(\mathsf{c}_1,\mathsf{c}_2;\big\{\mathsf{f}^{(j)}_\alpha\big\}\big)$ to vanishing values of the anisotropy scaling variables $\mathsf{f}^{(j)}_\alpha=f^{(j)}_\alpha L^{\Psi/\nu}$ is precisely the scaling function $D(\mathsf{c}_1,\mathsf{c}_2)$ determined to order $\epsilon$ in Ref.~\cite{SD08}. If we additionally set $\mathsf{c}_j$ to the  fixed-point values $0$ and $\infty$ of the isotropic special and ordinary transitions, we must recover the respective Casimir amplitudes $\Delta_C^{(\text{sp},\text{sp})}$, $\Delta_C^{(\text{sp},\text{ord})}$, and $\Delta_C^{(\text{ord},\text{ord})}$; i.e.,
\begin{equation}
\Delta_C^{(\wp)}=
\begin{cases}
D(0,0;\{0\})&\text{for }\wp=\text{sp},\text{sp},\\
D(\infty,\infty;\{0\})&\text{for } \wp=\text{ord},\text{ord},\\
D(0,\infty;\{0\})&\text{for }\wp=\text{sp},\text{ord}.
\end{cases}
\end{equation}
The expansions of $\Delta_C^{(\text{ord},\text{ord})}$, and of $\Delta_C^{(\text{sp},\text{sp})}$  and $\Delta_C^{(\text{sp},\text{ord})}$, were first determined to $O(\epsilon)$ in Refs.~\cite{Sym81}  and \cite{KD91}, respectively. That the small-$\epsilon$ expansion of $\Delta_C^{(\text{sp},\text{sp})}$ involves fractional powers of $\epsilon$ modulo logarithms beyond $O(\epsilon)$ was shown in our work with Shpot \cite{DGS06} and Ref.~\cite{GD08}.

To proceed we must first recall some background related to the anisotropic special transition in semi-infinite systems.

\subsection{Anisotropic special transitions}\label{sec:msp}

In order to characterize $m$-special transitions in semi-infinite systems, it will be helpful to introduce the (zero-field) surface susceptibilities
\begin{equation}
\label{eq:chi11def}
\int_{\mathfrak{B}_{j}}\mathrm{d} A\,G^{(0,2)}_{\alpha_1,\alpha_2}(\bm{X}_j,\bm{X}_k)
=\chi_{jk;\alpha_1}\,\delta_{\alpha_1\alpha_2}\,,
\end{equation}
where $\bm{X}_k$ is an arbitrary point on $\mathfrak{B}_k$. The function $\chi_{jk;\alpha}$ describes the response of $\phi_\alpha(\bm{X}_j)$ to a surface magnetic field $\bm{h}^{(k)}=h^{(k)}\bm{e}_\alpha$ localized on $\mathfrak{B}_k$ and oriented along the $\alpha$-axis. 

At an $m_1$-special transition of a semi-infinite system with surface plane $\mathfrak{B}_1$, $m_1$ of the $n$ enhancement variables $\mathring{c}_\alpha^{(1)}$ of this plane are equal and take a special value $\mathring{c}_{m_1\text{-sp}}$ while the remaining ones are larger, i.e., are subcritically enhanced relative to this value $\mathring{c}_{m_1\text{-sp}}$. Let us label the  $m_1$ ``easy axes''  spanning the subspace  in which the orientation of the order parameter is energetically favored on $\mathfrak{B}_1$ as $\alpha^{(1)}_e$ and the remaining ones pertaining to the local ``hard axes'' as $\alpha_h^{(1)}$. In order that an $m_1$-special transition occurs in the semi-infinite system with surface $\mathfrak{B}_1$ (in sufficiently high dimensions $d$), we must have
\begin{equation}
\label{eq:m1sp}
\begin{array}{ccc@{\;\;}c}
  \mathring{c}^{(1)}_\alpha
&=&\mathring{c}_{m_1\text{-sp}}&\text{for }\alpha\in\{\alpha_e^{(1)}\}\,,\\[\medskipamount]     
 \mathring{c}^{(1)}_\alpha & >&\mathring{c}_{m_1\text{-sp}}&\text{for }\alpha\in\{\alpha_h^{(1)}\}\,.
  \end{array}
\end{equation}

At such a transition, the local surface susceptibilities $\left.\chi_{11;\alpha}\right|_{L=\infty}$ have temperature singularities of the form
\begin{equation}
\label{eq:chi11msptau}
\left.\chi_{11;\alpha}\right|_{L=\infty}\sim
 \begin{cases}
 t^{-\gamma^{(1)}_{11,e}}&\text{for }\alpha\in\{\alpha_{e}^{(1)}\}\,,\\[\medskipamount]
C_{1,\alpha}+C_{2,\alpha}\, t^{-\gamma^{(1)}_{11,h}}&\text{for }\alpha\in\{\alpha_h^{(1)}\}\,.
 \end{cases}
\end{equation}
Here $C_{1,\alpha}$ and $C_{2,\alpha}$ are nonuniversal constants. The  $\epsilon$~expansions of the surface exponents $\gamma^{(1)}_{11,e}$ and $\gamma_{11,h}^{(1)}$ coincide with those given in Eqs.~(7) and (8) of Ref.~\cite{DE84} for the critical exponents denoted $\gamma_{11,e}$ and $\gamma_{11,h}$ there, where the parameters $m_h$ and $m_e$  must be set to $n-m_1$ and $m_1$, respectively.%
\footnote{In the case of $\gamma^{(2)}_{11,e}$ and $\gamma_{11,h}^{(2)}$, their analogs for semi-infinite systems with surface $\mathfrak{B}_2$, the surface susceptibilities $\chi_{22,\alpha}$ obviously take over the roles of $\chi_{11,\alpha}$ in Eq.~\eqref{eq:chi11msptau}, and one must set $m_e=m_2$ and $m_h=n-m_2$ in the cited $\epsilon$ expansions.} 
Since $\gamma_e^{(1)}>0$, while $\gamma_h^{(1)}<0$, the surface $\mathfrak{B}_1$ exhibits $O(m_1)$ surface criticality at the transition point in the sense that the $\chi_{11;\alpha}|_{L=\infty}$ diverge or have a cusp singularity depending on whether $\alpha=\alpha_e^{(1)}$ or $\alpha=\alpha_h^{(1)}$. That is, 
\begin{equation}
 \label{eq:charmsp}
\left.\chi_{11;\alpha}^{-1}\right|_{L=\infty,m_1\text{-sp}}=
 \begin{cases}0&\text{for }\alpha\in\{\alpha^{(1)}_e\},\\[\medskipamount]
\text{const} <\infty&\text{for }\alpha\in\{\alpha_h^{(1)}\},
 \end{cases}
\end{equation}
where the subscript $m_1\text{-sp}$ indicates that $t=0$ and the conditions~\eqref{eq:m1sp} are satisfied.

Turning to the renormalized surface susceptibility $\left.\chi_{11;\alpha,R}\right|_{L=\infty}$, we note that the RG equations~\eqref{eq:GRGiso} yield the scaling form
\begin{equation}
\label{eq:chi11renscf}
\left.\chi_{11;\alpha,R}\right|_{L=\infty}\approx\tau^{-\gamma_{11}^{sp}}\,X_\alpha\big(c^{(1)}\tau^{-\Phi},\{f_j\tau^{-\Psi}\}\big),
\end{equation}
where $\gamma_{11}^{sp}=\nu[1-\eta_\phi(u^*)-\eta_1(u^*)]$ is a familiar surface exponent of  the isotropic special transition whose $\epsilon$ expansion to order $\epsilon^2$ can be found in Eq.~(3.156f) of Ref.~\cite{Die86a}. Since $\tau\sim t$, consistency of the crossover scaling form with the asymptotic behavior~\eqref{eq:m1sp} at the $m_1\text{-sp}$ transition requires that the scaling functions $X_\alpha$ with $\alpha\in\{\alpha_e^{(1)}\}$ have  singularities. To locate the $m_1\text{-sp}$ transition, we set $c_\alpha^{(1)}\equiv c_{e}^{(1)}$ for all $\alpha\in\{\alpha^{(1)}_e\}$. 
Fulfillment of this condition guarantees that we have an $O(m_1)$ symmetry in the subspace of the $m_1$ easy-axes components $\alpha^{(1)}_e$ (provided this symmetry is not spontaneously broken). Thus, all scaling functions $X_\alpha$ with $\alpha\in\{\alpha^{(1)}_e\} $ become identical. For given values of the anisotropy variables $\{f^{(1)}_j\}$, we can determine the critical value $c_{m_1\text{-sp}}$ of $c^{(1)}$ associated with the $m_1\text{-sp}$ transition from the condition that $\lim_{t\to 0}1/\chi_{11,\alpha_e}|_{L=\infty}$ vanishes at $c^{(1)}=c_{m_1\text{-sp}}$ (while keeping the constraint $c_\alpha^{(1)}\equiv c_{e}^{(1)}$ for all $\alpha\in \{\alpha_e\}$). The shift $c_{m_1\text{-sp}}-c_{\text{sp}}$ must have the scaling form 
\begin{equation}
 \label{eq:shift}
c_{m_1\text{-sp}}-c_{\text{sp}}=\big[f^{(1)}_1\big]^{\Psi/\Phi}\,W(f^{(1)}_2/f^{(1)}_1,\dotsc,f^{(1)}_{n-1}/f^{(1)}_1)\,.
\end{equation}

To determine the temperature singularity of the surface susceptibility at the $m_1\text{-sp}$ transition from the singularity of its scaling function is a rather cumbersome procedure because the RG scheme used so far does not directly yield the power $\tau^{-\gamma_{11,e}^{(1)}}$ in exponentiated form. As is shown in Ref.~\cite{DE84}, the required information can be obtained in an easier and more direct fashion by formulating the RG against an anisotropic background rather than the $O(n)$~symmetric theory used above. Since the order parameter components $\phi_\beta$ with $\beta\in\{\beta^{(j)}_h\}$ are \emph{noncritical} on the boundary planes $\mathfrak{B}_j$ in the sense that the respective inverse surface susceptibilities $1/\chi_{jj,\beta}|_{L=\infty}$ do not vanish at the transition, we can set the associated bare enhancement variables $\mathring{c}^{(j)}_\beta$ with $\beta\in\{\beta_h^{(j)}\}$ to the fixed-point values $c^*_{\text{ord}}=\infty$ and expand about the $m_j\text{-sp}$ multicritical points of the corresponding semi-infinite systems with surfaces $\mathfrak{B}_j$. 

To this end we form, by analogy with Eq.~\eqref{eq:transf},  the symmetric combination
\begin{equation}
 \label{eq:cdot}
\mathring{c}_e^{(j)}=\frac{1}{m_j}\sum_{\alpha\in\{\alpha^{(j)}_e\}}\mathring{c}^{(j)}_\alpha
\end{equation}
and $m_j-1$ orthogonal anisotropy variables $\mathring{g}_k^{(j)}$, $k=1,\dotsc,m_j-1$, from the coefficients $\mathring{c}_\alpha^{(j)}$ with $\alpha\in\{\alpha^{(j)}_e\}$. The former couples to the $O(m_j)$~symmetric operator $\sum_{\alpha\in\{\alpha_e^{(j)}\}}\phi_\alpha^2$. The anisotropies $\mathring{g}_k^{(j)}$ couple to analogs of the anisotropy operators $A_\alpha$ for the respective $m_j$-dimensional subspaces, which we denote as $Q_{k}^{(j)}$. Let $Q_{k,R}^{(j)}=Z_\phi^{-1}Q_{k}^{(j)}$ be the operators one obtains through the substitutions $\phi_\alpha^2\to \phi_{\alpha,R}^2=Z_\phi^{-1}\phi_\alpha^2$ for all $\alpha\in\{\alpha^{(j)}_e\}$. Following Ref.~\cite{DE84}, we can introduce renormalized variables $c_e^{(j)}$ and $g_k^{(j)}$ by writing the  boundary contributions to the action as
\begin{eqnarray}
 \label{eq:L1expmsp}
\int_{\partial\mathfrak{V}}\mathcal{L}_{\partial\mathfrak{V}}[Z_\phi\bm{\phi}_R]\,\mathrm{d}{A}&=&\bigg[(\mu Z_{e}^{(j)}c_e^{(j)}+\mathring{c}_{m_1\text{-sp}})Z_\phi\sum_{\alpha\in\{\alpha_e^{(j)}\}}\phi_{\alpha,R}^2\nonumber\\ &&\strut
 + \mu\,Z^{(j)}_{g}Z_\phi\sum_kg^{(j)}_kQ^{(j)}_{k,R}+c^*_{\text{ord}}\,Z_\phi\sum_{\alpha\in\{\alpha^{(j)}_h\}}\phi_{\alpha,R}^2\bigg].
\end{eqnarray}
Here $Z^{(j)}_{e}$ and $Z^{(j)}_g$ are respectively the renormalization factors  $Z_{\dot{c}}$ and $Z_{\dot{f}}$  given in  Eqs.~(70) and(71) of this reference, with $m_e=m_j$ and $m_h= n-m_j$. The superscripts $^{(j)}$  here serve to remind us that these functions depend via $m_j$ on the number $j$ of the boundary plane $\mathfrak{B}_j$.  Just as $\mathring{c}_{\text{sp}}$, the critical value $\mathring{c}_{m_1\text{-sp}}$ vanishes in a perturbative approach in $4-\epsilon$ dimensions if dimensional regularization is used. (In a theory regularized by means of a large-momentum cutoff $\Lambda$, both quantities would diverge $\sim \Lambda$ in $d=4$ dimensions.)

The RG eigenvalues  governing the flow of the running variables $\bar{c}^{(j)}_e(\ell)\sim \ell^{-y^{(j)}_e}c_e^{(j)}$ and $\bar{g}^{(j)}_k(\ell)\sim\ell^{-y^{(j)}_g}g^{(j)}_k$ are given by
\begin{eqnarray}
y_e^{(j)}&=&1+\eta_e^{(j)}(u^*), \nonumber\\ y^{(j)}_g&=&1+\eta^{(j)}_g(u^*),
\end{eqnarray}
where the exponent functions are defined as
\begin{equation}
 \label{eq:etaeg}
\eta^{(j)}_{e,g}(u)\equiv \mu\partial_\mu|_0\ln Z_{e,g}\,,
\end{equation}
by analogy with Eq.~\eqref{eq:betaexpfctns}. The $\epsilon$~expansions of $\nu y_e^{(j)}$ and $\nu y_g^{(j)}$ may be found from those given  in Eqs.~(87) and (88) of Ref.~\cite{DE84} (for the exponents denoted $\dot{\Phi}$ and $\dot{\Psi}$ there) by setting the parameter $m_h$ to $n-m_j$.

It should be clear that the generalized asymptotic boundary conditions with which we are concerned --- namely, those for which $m_j$ order parameter components $\phi_\alpha$, $\alpha\in\{\alpha_e^{(j)}\}$, are critically enhanced on $\mathfrak{B}_j$ while the remaining $n-m_j$ ones with $\alpha\in\{\alpha_h^{(j)}\}$ are subcritically enhanced there ---
are associated with fixed points located at $\big(c_e^{(1)},c_e^{(2)}\big)=(0,0)$ on the zero-anisotropy hyperplane $g^{(j)}_k=0$ with $\tau=0$ and $u=u^*$. These fixed points are unstable in the $c_e^{(j)}$ and $g^{(j)}_k$ directions. By contrast, surface spin anisotropies on $\mathfrak{B}_j$ that are restricted to the respective subspaces of hard axes $\alpha\in\{\alpha^{(j)}_h\}$ are irrelevant. Components $\phi_\alpha$ whose index $\alpha$ belongs to the set of hard axes, $\{\alpha_h^{(j)}\}$, satisfy Dirichlet boundary conditions at the plane $\mathfrak{B}_j$. The behavior near the plane can be obtained from the short-distance expansion (cf.\ Ref.~\cite{DD81b} and \cite[p.\ 190ff]{Die86a})
\begin{equation}
 \label{eq:SDE}
\phi_{\beta,R}(\bm{X}+\bm{n}\Delta z)\mathop{\approx}\limits_{\Delta z\to 0}C(\Delta z)\,[\phi_{\beta}(\bm{X})]_R\,,
\end{equation}
where the operator on the right-hand side means the renormalized surface operator $[\phi_{\beta}(\bm{X})]_R=(Z_\phi Z_{1,e})^{-1/2}\phi_\beta(\bm{X})$ at a point $\bm{X}\in\mathfrak{B}_j$. The behavior of the function $C(\Delta z)$ at distances $\delta z$ small compared to $L$ and $\xi_\infty$ (but large compared to microscopic distances) is governed by the difference of scaling dimensions of the operators on the left-hand and right-hand sides of Eq.~\eqref{eq:SDE}. We have
\begin{equation}
 \label{eq:C}
C(\Delta z)\mathop{\approx}\limits_{\Delta z\to 0}\mbox{const}\, |\Delta z|^{\sigma_e^{(j)}}\quad\mbox{with } \sigma_e(n,m_j,d)=(\eta_{\|,e}^{(j)}-\eta)/2,
\end{equation}
where $\eta_{\|,e}^{(j)}$ is the surface critical  exponent $\eta_{\|,e}$ of Ref.~\cite{DE84} for $m_h=n-m_j$. Using  the results obtained there, one arrives at the series expansions
\begin{eqnarray}\label{eq:sigexp}
\sigma_e(m_j,n,d)&=&\frac{n-2 m_j-2}{6}\,u^* +
\frac{6+5m_j-2n}{9}\, {(u^*)}^{2}+O\big[{(u^*)}^3\big] \nonumber\\
&=&\frac{n-2m_j-2}{2(8+n)}\,\epsilon +\frac{12+(4+5n)n-4(2n+1)m_j}{2(8+n)^3}\,\epsilon^2+O(\epsilon^3).\nonumber\\
\end{eqnarray}
Of interest is the case of a $(d=3)$-dimensional Heisenberg magnet with an easy-axis spin anisotropy at $\mathfrak{B}_j$ (i.e., $d=n=3, m_j=1$).  To obtain a rough estimate of the exponent $\sigma_e(3,1,3)$, we evaluate the $O(\epsilon^2)$ expression~\eqref{eq:sigexp} at $\epsilon=1$. This yields  $\sigma_e(3,1,3)\simeq -40/1331\simeq-0.03$. It is plausible that a small negative value results.%
\footnote{For the analogous exponent $(\eta_{\|}^{\text{sp}}-\eta)/2$ of the $(d=3)$-dimensional semi-infinite Ising model's special transition, the results $y_h=2.482$ and $y^{\text{sp}}_{h_1}=1.636$ given in Ref.~\cite{DBN05} for the RG eigenvalues $y_h$ and $y^{\text{sp}}_{h_1}$ of the bulk and surface magnetic fields yield $(\eta_{\|}^{\text{sp}}-\eta)/2= y_h-1-y^{\text{sp}}_{h_1}\simeq-0.154$.} 
However, because of our na\"{\i}ve extrapolation procedure, this estimate  is not very reliable. Unfortunately, we are not aware of accurate Monte Carlo estimates of this exponent. Although a number of detailed Monte Carlo investigations  of $O(n)$ models on simple cubic lattices \cite{Kre00,DKK02,DBN05} were carried out during the past decade, none of these allowed for $O(n)$-symmetry breaking surface spin anisotropies.

\section{Casimir amplitudes}\label{sec:CA}
\subsection{Calculation of residual free energy via RG improved perturbation theory}
\label{sec:casicalc}
We are now ready to turn to the calculation of  Casimir amplitudes. The general case of asymptotic boundary conditions we wish to consider should be clear from the foregoing section: those for which the renormalized enhancement and anisotropy variables $c_e^{(j)}$ and $g^{(j)}_k$ of boundary plane $\mathfrak{B}_j$ take their fixed point values $c_e^{(j)}=g^{(j)}_k=0$ while all  bare ``hard-axes'' surface enhancement variables $\mathring{c}^{(j)}_\alpha$ with $\alpha\in\{\alpha_h^{(j)}\}$ are set to the value $c^*_{\text{ord}}=\infty$. The conditions $c_e^{(j)}=g^{(j)}_k=0$ imply that all renormalized easy-axes enhancement variables $c_\alpha^{(j)}$ with $\alpha\in\{\alpha_e^{(j)}\}$ vanish. In general, we therefore have: (i)
 $m_{c,c}$ of components $\phi_\alpha$   that are critically enhanced on both planes ($\alpha\in\{\alpha^{(1)}_e\}\cap\{\alpha^{(2)}_e\}$); (ii) $m_{\mathrm{D},\mathrm{D}}$ components satisfying  Dirichlet boundary conditions at both planes; (iii) $m_{c,\mathrm{D}}$ and $m_{\mathrm{D},c}$ components that are critically enhanced on $\mathfrak{B}_1$ and $\mathfrak{B}_2$ but satisfy Dirichlet boundary conditions on $\mathfrak{B}_2$ and $\mathfrak{B}_1$, respectively. Components that are critically enhanced at $\mathfrak{B}_j$ have an asymptotic near-boundary behavior of the form specified by Eqs.~\eqref{eq:SDE}--\eqref{eq:sigexp} near this plane, irrespective of whether they are critically enhanced at the complementary plane or satisfy Dirichlet boundary conditions there.

To determine the corresponding Casimir amplitudes $\Delta_C$ via RG improved perturbation theory in $d=4-\epsilon$ dimensions, we set all enhancement variables to the above-mentioned fixed-point values and compute the residual free energy~\eqref{eq:frescrit} at the bulk critical point. Whenever $m_{c,c}>0$,  there are free propagators $G_{L,\alpha}=G_L^{(\mathrm{N,N})}$ among those of Eq.~\eqref{eq:Gbc} whose spectral decompositions\eqref{eq:Gfreespecrep} involve the eigenfunction $\varphi^{(\mathrm{N,N)}}_{r=0}$. Since the associated eigenvalue vanishes, we encounter a zero-mode problem of the kind dealt with in Refs.~\cite{DGS06} and \cite{GD08}. Its origin is that Landau theory in those cases erroneously predicts  a sharp transition at $T_{c,\infty}$ for films of finite thickness $L$. The infrared singularities associated with these zero modes entail that conventional RG improved perturbation theory becomes ill-defined at $T_{c,\infty}$. As shown in Refs.~\cite{DGS06} and \cite{GD08}, this can be remedied by a reorganization of RG improved perturbation theory. The strategy is analogous to the one used in the theory of finite size effects on continuous phase transitions in systems that are finite in all, or in all but one, dimensions \cite{BZJ85,RGJ85}. Its crux is to split off the zero-mode components from the $m_{c,c}$ fields $\phi_\alpha(\bm{x})$ with $\alpha\in\{\alpha^{(1)}_e\}\cap\{\alpha^{(2)}_e\}$ and integrate out the remaining degrees of freedom to construct an effective action for the  zero-mode components. To this end we introduce the $m_{c,c}$-component field $\bm{\varphi}(\bm{y})$ with components 
\begin{equation}
 \label{eq:varphi}
 \varphi_\alpha(\bm{y})=L^{-1/2}\int_0^L \mathrm{d} z\,\phi_\alpha(\bm{y},z)\,,\quad\alpha\in \{\alpha^{(1)}_e\}\cap\{\alpha^{(2)}_e\},
\end{equation}
and decompose $\bm{\phi}(\bm{x})$ as
\begin{equation}
\label{eq:splitphi}
\phi_\alpha(\bm{y},z)=
\begin{cases}
L^{-1/2}\varphi_\alpha(\bm{y})+\psi_\alpha(\bm{y},z)&\text{for }\alpha\in\{\alpha^{(1)}_e\}\cap\{\alpha^{(2)}_e\},\\
\psi_\alpha(\bm{y},z)&\text{for }\alpha\notin\{\alpha^{(1)}_e\}\cap\{\alpha^{(2)}_e\}.
\end{cases}
\end{equation}
Obviously, the field $\bm{\varphi}$ contains the zero-mode contributions to $\bm{\phi}$. By orthogonality to the $r\ne 0$ eigenfunctions $\varphi_r^{(\text{N,N})}$, we therefore have
\begin{equation}
 \label{eq:psizero}
 \int_0^L\mathrm{d} z\,\psi_\alpha(\bm{y},z)=0 \quad\text{for all }\alpha\in\{\alpha^{(1)}_e\}\cap\{\alpha^{(2)}_e\}.
\end{equation}

Let us introduce the free energy associated with the field $\psi$, $F_\psi$,  and the corresponding reduced area density $f_\psi$ by
\begin{equation}
\label{eq:Fpsi}
\frac{F_{\psi}}{k_BT}=-\ln\text{Tr}_{\psi}\mathrm{e}^{-\mathcal{H}[\bm{\psi}]},\quad f_\psi(L)\equiv\lim_{A\to\infty}\frac{F_{\psi}}{k_BTA},
\end{equation}
where $\text{Tr}_\psi(.)$ means the functional integral $\int \mathcal{D}\bm{\psi}(.)$. When $m_{c,c}=0$, no  zero modes occur in the spectral decomposition of the free propagators at $T_{c,\infty}$; then $F_\psi$ and $f_\psi(L)$ coincide with the total free energy $F$ and the reduced area density $f_L$, respectively. Whenever $m_{c,c}>0$, the effective Hamiltonian, defined by
\begin{equation}
\label{eq:Heff}
\mathrm{e}^{-\mathcal{H}_{\text{eff}}[\bm{\varphi}]}=\mathrm{e}^{F_\psi/k_BT}\,\text{Tr}_\psi\mathrm{e}^{-\mathcal{H}[\bm{\phi}[\bm{\varphi},\bm{\psi}]]},
\end{equation}
is nonzero and yields the additional contribution
\begin{equation}
\label{eq:Fvarphi}
f_\varphi(L)=\lim_{A\to\infty} \frac{F_{\varphi}}{k_BTA}=-\lim_{A\to\infty}\frac{1}{A}\ln\text{Tr}_{\varphi}\mathrm{e}^{-\mathcal{H}_{\text{eff}}[\bm{\varphi}]}
\end{equation}
to the reduced area density
\begin{equation}
 \label{eq:fLfpsiFvarphi}
f_{L}=f_{\psi}(L)+f_{\varphi}(L).
\end{equation}

To compute $f_\psi(L)$, we can use a standard loop expansion
\begin{equation}
-f_{\psi}(L)A=\,\,\raisebox{-8pt}{\includegraphics[width=22pt,clip]{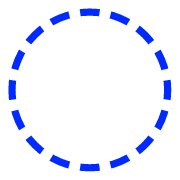}}\,\,+\,\,\raisebox{-14.75pt}{\includegraphics[width=18pt,clip]{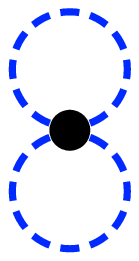}}\,\,+O(3\text{-loops}).
\end{equation}
Here the dashed blue lines in the second graph represent free $\psi$-propagators $G_{\psi,L,\alpha}=[Q_0(-\Delta^{(\mathrm{A}_\alpha,\mathrm{B}_\alpha)}+\mathring{\tau})\delta_{\alpha\beta}Q_0]^{-1}$, where  $\Delta^{(\mathrm{A}_\alpha,\mathrm{B}_\alpha)}$ with $\mathrm{A}_\alpha,\mathrm{B}_\alpha=\mathrm{N},\mathrm{D}$ means the Laplacian, subject to the boundary conditions $(\mathrm{A}_\alpha,\mathrm{B}_\alpha)$ applying to this value of $\alpha$. Further, the operator $Q_0$ projects onto the subspace orthogonal to the $k_0=0$ modes. In the $\bm{p}z$~representation we have
\begin{equation}
\label{eq:GpsiL}
G_{\psi,L,\alpha}(\bm{p};z_1,z_2)=G^{(\mathrm{A}_\alpha,\mathrm{B}_\alpha)}_{L}(\bm{p};z_1,z_2)-\delta_{\mathrm{A}_\alpha,\mathrm{N}}\,\delta_{\mathrm{B}_\alpha,\mathrm{N}}
\frac{1}{L(p^2+\mathring{\tau})},
\end{equation}
as follows from Eqs.~\eqref{eq:Gfreespecrep} and \eqref{eq:varphiAB}.

The one-loop contribution
\begin{equation}
 \label{eq:fpsioneloopgr}
\raisebox{-8pt}{\includegraphics[width=22pt,clip]{./oneloopfpsi.eps}}=\frac{1}{2}\text{Tr}\ln[Q_0(-\Delta^{(\mathrm{A}_\alpha,\mathrm{B}_\alpha)}+\mathring{\tau})\delta_{\alpha\beta}Q_0]
\end{equation}
splits into a sum $\sum_\alpha$ of contributions which may be gleaned from the results for Casimir amplitudes of the Gaussian theory under the boundary conditions $(\text{A}_\alpha,\text{B}_\alpha)$,  with $\text{A}_\alpha,\text{B}_\alpha=\text{D},\text{N}$, given in Refs.~\cite{Sym81}, \cite{KD91}, and \cite{KD92a}.
The resulting one-loop contribution to the residual free energy density $f_{\psi,\text{res}}|_{T=T_{c,\infty}}$ becomes
\begin{equation}
\label{eq:fpsiresoneloop}
f^{[1]}_{\psi,\text{res}}(L)\big|_{T=T_{c,\infty}}=-\frac{\Gamma(d/2)\,\zeta(d)}{2^d\,\pi^{d/2}L^{d-1}}\,\big[m_{\mathrm{D},\mathrm{D}}+m_{c,c}+(2^{1-d}-1)(m_{c,\mathrm{D}}+m_{\mathrm{D},c})\big],
\end{equation}
where $\zeta(d)$ is the Riemann zeta function.

The two-loop contribution to $f_\psi(L)$ can be expressed in terms of the integrals
\begin{equation}
 \label{eq:Iwp2}
I_{2}^{(\mathrm{A}_\alpha,\mathrm{B}_\alpha;\mathrm{A}_\beta,\mathrm{B}_\beta)}(L;\mathring{\tau})\equiv\int_{0}^{L}\frac{\mathrm{d}{z}}{L}\,G_{L,\psi,\alpha}(\bm{x};\bm{x})\, G_{L,\psi,\beta}(\bm{x};\bm{x})
\end{equation}
as
 \begin{equation}
 \label{eq:fpsi2gr}
f_{\psi}^{[2]}(L)  = \strut -A^{-1}\,\,\raisebox{-12pt}{\includegraphics[width=16pt,clip]{./twoloopfpsi.eps}} \,\strut
  = \frac{\mathring{u}L}{4!}
   \sum_{\alpha,\beta}I_{2}^{(\mathrm{A}_\alpha,\mathrm{B}_\alpha;\mathrm{A}_\beta,\mathrm{B}_\beta)}(L;\mathring{\tau})(1+2\delta_{\alpha\beta}).
 \end{equation}
In previous calculations \cite{KD91,KD92a,DGS06,GD08}, only the special cases of  the integrals~\eqref{eq:Iwp2} with $\alpha=\beta$ were encountered. We need these integrals  at $\mathring{\tau}=0$ also for $\alpha\ne \beta$. Their $\epsilon$~expansion to first order is worked out in Appendix~\ref{app:requint}. Upon inserting the results into Eq.~\eqref{eq:fpsi2gr} and expressing $\mathring{u}$ in terms of the renormalized coupling constant $u$, we obtain
\begin{eqnarray}\label{eq:fpsirestwoloop}
f_{\psi,\text{res}}^{[2]}(L)\big|_{T=T_{c,\infty}}
 & = & L^{-(d-1)}\,\frac{\pi^{2}u[1+O(u)]}{4608}\Big[\big(m_{c,c}+m_{\mathrm{D,D}}\big)\big(2+3m_{c,c}+3m_{\mathrm{D,D}}\big) \nonumber \\ &&\strut +\frac{n}{3}\big(n+2-6m_{c,c}-6m_{\mathrm{D,D}}\big)
 +O(\epsilon)\Big].
 \end{eqnarray}

We turn next to the calculation of the zero-mode contribution $f_{\varphi,\text{res}}$. The one-loop graph
\raisebox{-0.5em}{\includegraphics[width=30pt]{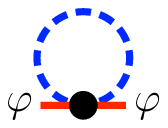}} yields to $\mathcal{H}_{\text{eff}}[\varphi]$ a contribution quadratic in $\varphi$ which changes the coefficient of $\varphi^2/2$ from $\mathring{\tau}$ to 
\begin{eqnarray}
 \label{eq:baretauL}
\mathring{\tau}_{L}&=&\mathring{\tau}+\frac{\mathring{u}}{3!}\int_{0}^{L}\frac{\mathrm{d}z}{L}\,\Big[2G^{\text{(N,N)}}(\bm{x};\bm{x})+\sum_{\alpha=1}^nG_{L,\psi,\alpha}(\bm{x};\bm{x})\Big] \nonumber\\
&=&\mathring{\tau}+\frac{\mathring{u}}{3!}\big[(m_{c,c}+2)\,I_1^{\text{(N,N)}}(L;\mathring{\tau}) +m_{\mathrm{D},\mathrm{D}}\,I_1^{\text{(D,D)}}(L;\mathring{\tau}) \nonumber\\ &&\strut \phantom{\mathring{\tau}+\frac{\mathring{u}}{3!}\big[}
+(m_{c,\mathrm{D}}+m_{\mathrm{D},c})\,I_1^{\text{(N,D)}}(L;\mathring{\tau})\big].
\end{eqnarray}

The required integrals 
\begin{equation}
 \label{eq:I1def}
 I^{\text{(A,B)}}_1(L;\mathring{\tau})=\int_0^L\frac{\mathrm{d}{z}}{L}\,G^{\text{A,B)}}_{L,\psi}(\bm{x};\bm{x})\,,\;\;\mathrm{A}, \mathrm{B}=\mathrm{N},\mathrm{D},
\end{equation}
with $\mathring{\tau}=0$ are computed in Appendix~\ref{app:requint}. The results are
\begin{eqnarray}
 \label{eq:I1res}
 I^{\text{(A,B)}}_1(L;0)=-\frac{2^{1-d} \pi ^{(d-3)/2}  \zeta (3-d) }{\Gamma[(d-1)/2)]\cos
   (\pi  d/2)\,L^{d-2}}\times
   \begin{cases}
   1&\text{for }\mathrm{A}=\mathrm{B},\\
   (2^{3-d}-1)&\text{for }\mathrm{A}\ne\mathrm{B}.
   \end{cases} \nonumber\\
 \end{eqnarray}
Substituting them into Eq.~\eqref{eq:baretauL}, expressing $\mathring{u}$ in terms of the renormalized coupling constant $u$, and expanding in $\epsilon$ then gives
\begin{equation}
\label{eq:tauL}
\mathring{\tau}_L|_{T=T_{c,\infty}}=\frac{u}{L^{2}}\,[1+O(u)]\frac{\pi^{2}}{36}\,[3m_{c,c}+3m_{\mathrm{D,D}}-n+4+O(\epsilon)].
\end{equation}

The contribution from the zero mode 
\begin{equation}
 \label{eq:Fgexpfvarphi}
A\,f_{\varphi}(L)=\,-\,\raisebox{-7pt}{\includegraphics[width=20pt,clip]{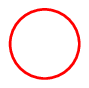}}\,\,-\,\,\raisebox{-12pt}{\includegraphics[width=16pt,clip]{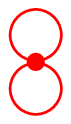}}\,\,+\ldots.
\end{equation}
can now be easily inferred from the results of Ref.~\cite{GD08}. To obtain the one-loop contribution to the residual free energy density $f_{\varphi,\text{res}}(L)$ we must simply set $n=m_{c,c}$ in its Eq.~(4.28) and replace the variable $\mathring{r}_L$ it involves by the quantity $\mathring{\tau}_L$ given in Eq.~\eqref{eq:tauL}. The two-loop term follows in the same manner, except that an additional sign change must be made (as can be seen from Eqs.~(4.23) and (4.27) of Ref.~\cite{GD08}). Hence we find
\begin{equation}
 \label{eq:fvarphitwoloop}
f_{\varphi,\text{res}}(L)\big|_{T=T_{c,\infty}}  =  -m_{c,c}\,\frac{A_{d-1}}{d-1}\,\mathring{\tau}_{L}^{(d-1)/2}+\frac{\mu^\epsilon u}{ L}\,\frac{m_{c,c}(m_{c,c}+2)}{4!}\, \frac{A_{d-1}^{2}}{N_d}\,\mathring{\tau}_{L}^{d-3}+\ldots
\end{equation}
with 
\begin{equation}
\label{eq:Ad}
A_{d}=-(4\pi)^{-d/2}\Gamma(1-d/2)\,,
\end{equation}
where the $\mathring{\tau}_L$ on the right-hand side of Eq.~\eqref{eq:fvarphitwoloop} are to be taken at the bulk critical point. Since $\mathring{\tau}_L|_{T_{c,\infty}}$ is linear in $u$, its value at $u=u^*$ is of order $\epsilon$. This implies that the first term in Eq.~\eqref{eq:fvarphitwoloop} yields a contribution of order $\epsilon^{3/2}$ to the Casimir amplitude. The contribution from the second term is of order $\epsilon^{2-\epsilon}$ and hence negligible to the order of our calculation ($\epsilon^{3/2}$).

\subsection{Results}

Before using the results of the foregoing section to determine the small-$\epsilon$ expansion of the Casimir amplitudes, we first give the results for the Gaussian ($\mathring{u}=0$) critical  ($\mathring{\tau}=0$) theory in general dimensions $d\ge 1$. The corresponding Casimir amplitudes $\Delta_{C,\mathrm{Gauss}}$ for the cases that $m_{\mathrm{D},\mathrm{D}}$  and $m_{\mathrm{N},\mathrm{N}}$   components $\phi_\alpha$ satisfy respectively Dirichlet  and  Neumann boundary conditions on both planes $\mathfrak{B}_j$, while the remaining $n-m_{\mathrm{D},\mathrm{D}}-m_{\mathrm{N},\mathrm{N}}=m_{\mathrm{D},\mathrm{N}}+m_{\mathrm{N},\mathrm{D}}$ components satisfy Dirichlet boundary conditions on one plane but Neumann boundary conditions on the complementary one, can be inferred from equation~\eqref{eq:fpsiresoneloop}. The reason is that in this Gaussian case, we do not have to worry about  the zero-mode contributions. Upon considering them for $
\mathring{\tau}>0$ [i.e.,  the first term in equation~\eqref{eq:fvarphitwoloop}] and then taking the limit $\mathring{\tau}\to 0+$, one sees that their contributions to $f_{\mathrm{res}}(L)$ vanish at $T=T_{c,\infty}$. Thus the Gaussian Casimir amplitudes are given by
\begin{equation}
 \label{eq:DeltaGauss}
\Delta_{C,\mathrm{Gauss}}=-\frac{\Gamma(d/2)\,\zeta(d)}{2^d\,\pi^{d/2}}\,\big[m_{\mathrm{D},\mathrm{D}}+m_{\mathrm{N},\mathrm{N}}+(2^{1-d}-1)(m_{\mathrm{N},\mathrm{D}}+m_{\mathrm{D},\mathrm{N}})\big].
\end{equation}

Returning to the interacting case $\mathring{u}\ne 0$, we
can add the results presented in Eqs.~\eqref{eq:fpsiresoneloop}, \eqref{eq:fpsirestwoloop}, and \eqref{eq:fvarphitwoloop} to obtain the renormalized residual free energy $f_{\text{res},R}(L,T,u)$. Its value at $u=u^*=3\epsilon/(n+8)+O(\epsilon^2)$ gives us the Casimir amplitude 
\begin{equation}
 \label{eq:DeltafresR}
\Delta_C=L^{d-1}\,f_{\text{res},R}(L,T_{c,\infty},u^*)\,.
\end{equation}
Its small-$\epsilon$ expansion becomes
\begin{equation}
 \label{eq:DeltaCexp}
 \Delta_C=a_0+a_1\epsilon+a_{3/2}\,\epsilon^{3/2}+o(\epsilon^{3/2})
 \end{equation}
 with
   \begin{equation}
 \label{eq:a0}
 a_0=-\frac{\pi^2}{768}\bigg(m_{c,c}+m_{\mathrm{D},\mathrm{D}}-\frac{7n}{15}\bigg),
 \end{equation}
 \begin{eqnarray}
  \label{eq:a1}
 a_1&=&-\frac{\pi^2}{1536}\bigg[\bigg(m_{c,c}+m_{\mathrm{D},\mathrm{D}}-\frac{7n}{15}\bigg)\bigg(\gamma_E-1+\ln(\pi)+\frac{28\ln 2}{15}-\frac{2\zeta'(4)}{\zeta(4)}\bigg)\nonumber\\ &&\phantom{-\frac{\pi^2}{1536}\bigg[\bigg(}\strut+\frac{16\,n\ln 2}{225}-\frac{(m_{c,c}+m_{\mathrm{D},\mathrm{D}})(2+3 m_{c,c}+3m_{\mathrm{D},\mathrm{D}})}{n+8}\nonumber\\ &&\phantom{-\frac{\pi^2}{1536}\bigg[\bigg(+\frac{16\,n\ln 2}{225}} \strut\strut-\frac{n(n+2-6m_{c,c}-6m_{\mathrm{D},\mathrm{D}})}{3(n+8)}\bigg],
\end{eqnarray}
and
\begin{equation}
  \label{eq:a3ov2}
 a_{3/2}=-\frac{m_{c,c}\,\pi^2}{288\sqrt{3}}\bigg(\frac{3m_{c,c}+3m_{\mathrm{D},\mathrm{D}}-n+4}{n+8}\bigg)^{3/2},
\end{equation}
where $\gamma_E=-\Gamma'(1)$ is the Euler-Mascheroni constant.

These expansions generalize previous results for critical Casimir amplitudes for those cases in which the $O(n)$ symmetry is not broken, neither spontaneously nor explicitly nor by boundary conditions, by the inclusion of symmetry breaking quadratic boundary terms.

Note that our results~\eqref{eq:DeltaCexp}--\eqref{eq:a3ov2} depend only on the sum of $m_{c,c}+m_{\mathrm{D},\mathrm{D}}$. Since $n-m_{c,c}-m_{\mathrm{D},\mathrm{D}}=m_{c,\mathrm{D}}+m_{\mathrm{D},c}$, the expansions of $\Delta_C$ to the given order of $\epsilon^{3/2}$ are insensitive to changes of $m_{c,\mathrm{D}}$ and $m_{\mathrm{D},c}$ at fixed $n$, $m_{c,c}$, and $m_{\mathrm{D},\mathrm{D}}$ that preserve the total number $ m_{\text{ul}}\equiv m_{c,\mathrm{D}}+m_{\mathrm{D},c}$ of components satisfying ``unlike'' boundary conditions, namely, Dirichlet boundary conditions on one boundary plane and those corresponding to critical enhancement on the complementary surface plane. The reason for this insensitivity is that the one-loop and two-loop graphs contributing to the given order are traces and local expressions in position space, respectively. The full theory does not have this property and hence in general will give different results for cases of different $m_{c,\mathrm{D}}$ with the same values of $n$, $m_{\mathrm{D},\mathrm{D}}$, and $m_{\text{ul}}$.  The obvious reason is that such cases with different values of $m_{c,\mathrm{D}}$ are physically distinct. As an illustrative example, one can compare
the cases of $(m_{c,c},m_{\mathrm{D},\mathrm{D}},m_{c,\mathrm{D}}, m_{\mathrm{D},c},n)=(0,1,2,0,3)$ and $(0,1,1,1,3)$ involving critically enhanced easy-plane interactions on $\mathfrak{B}_1$ and critically enhanced easy-axes interactions on both boundary planes, respectively.%
\footnote{For reasons discussed already in the Introduction, the $2$-special surface transition one is concerned with in the first case is expected to occur only when $d>3$. By contrast, the $1$-special transitions one is dealing with in the second case should be possible in $d=3$ dimension.} 

The special cases of $m_{\mathrm{D},\mathrm{D}}=n$ (Dirichlet boundary conditions on both planes), $m_{c,\mathrm{D}}+m_{\mathrm{D},c}=n$ (all components $\phi_\alpha$ satisfy Dirichlet boundary conditions on one boundary plane and are critically enhanced at the complementary one), and $m_{c,c}=n$ (all components are critically enhanced at both boundary planes) provide important checks of our results~\eqref{eq:DeltaCexp}--\eqref{eq:a3ov2}. In the first two cases, we indeed recover the series expansions
\begin{eqnarray}
\Delta_C(m_{\mathrm{D},\mathrm{D}}=n)/n&=&-\frac{\pi^{2}}{1440}+\epsilon\,\frac{\pi^{2}}{2880}\!\bigg[1-\gamma_E-\ln(4\pi)\nonumber\\ &&\strut +\frac{2\zeta^{\prime}(4)}{\zeta(4)}+\frac{5}{2}\,\frac{n+2}{n+8}\bigg]+O(\epsilon^{2}),
\end{eqnarray}
and
\begin{eqnarray}
\Delta_C(m_{c,\mathrm{D}}+m_{\mathrm{D},c}=n)/n&=&\frac{7\pi^{2}}{11520}-\epsilon\frac{7\pi^{2}}{23040}\bigg[1-\gamma_E-\frac{12\ln(2)}{7}\nonumber \\ && \strut -\ln(\pi) +\frac{2\zeta^{\prime}(4)}{\zeta(4)}-\frac{5}{7}\,\frac{n+2}{n+8}\bigg]+O(\epsilon^{2}),
\end{eqnarray}
of Krech and Dietrich \cite{KD91,KD92a} for the amplitudes $\Delta_{\mathrm{D},\mathrm{D}}/n$ and $\Delta_{\mathrm{D},\mathrm{sp}}/n$. Likewise in the third case, the resulting series expansion
\begin{eqnarray}
 \label{eq:Deltaepsexp}
\Delta_C(m_{c,c}=n)/n & = & -\frac{\pi^{2}}{1440}+\epsilon\frac{\pi^{2}}{2880}\!\left[1-\gamma_E-\ln(4\pi)+\frac{2\zeta^{\prime}(4)}{\zeta(4)}+\frac{5}{2}\,\frac{n+2}{n+8}\right]\nonumber \\ \strut &&\strut-\epsilon^{3/2}\frac{\pi^{2}}{72\sqrt{6}}\!\left(\frac{n+2}{n+8}\right)^{3/2}
 +O(\epsilon^{2})
 \end{eqnarray}
 agrees with the one obtained for $\Delta_{\mathrm{sp},\mathrm{sp}}/n$ in Ref.~\cite{DGS06}.
 
For ($d=3$)-dimensional Heisenberg systems ($n=3$), the following easy-axes cases are of interest:
\begin{eqnarray}
&&\text{\underline{(i) $m_{c,c}=1$, $m_{\mathrm{D},\mathrm{D}}=2$, $m_{c,\mathrm{D}}=0,m_{\mathrm{D},c}=0$, $n=3$:}}\nonumber\\
\label{eq:1,2,3}
&&\quad\Delta_C/3=
-0.00685-0.00377\, \epsilon 
-0.00572\, \epsilon ^{3/2}+o(\epsilon^{3/2}),\\[\bigskipamount]
&&\text{\underline{(ii) $m_{c,c}=0$, $m_{\mathrm{D},\mathrm{D}}=2$, $m_{c,\mathrm{D}}=1,m_{\mathrm{D},c}=0$, $n=3$:}}\nonumber\\
\label{eq:0,2,3}
&&\quad\Delta_C/3=-0.00257 - 0.00132\,\epsilon+O(\epsilon^2),\\[\bigskipamount]
&&\text{\underline{(iii) $m_{c,c}=0$, $m_{\mathrm{D},\mathrm{D}}=1$, $m_{c,\mathrm{D}}=1,m_{\mathrm{D},c}=1$, $n=3$:}}\nonumber\\
\label{eq:0,1,3}
&&\quad\Delta_C/3=0.00171 + 0.00230 \,\epsilon+O(\epsilon^2).
\end{eqnarray}

In Table~\ref{tab:Deltaest}, estimates for $\Delta_C/n$ at $d=3$ are given for  various choices of $m_{c,c}$, $m_{\mathrm{D},\mathrm{D}}$, $m_{c,\mathrm{D}}$, $m_{\mathrm{D},c}$, and $n$. In the columns labeled $O(\epsilon^\sigma)$ with $\sigma=0,1,3/2$ we have listed the values one obtains from the small-$\epsilon$  expansions~\eqref{eq:DeltaCexp} by truncating them at this order and evaluating them at $\epsilon=1$. The estimates shown in the column marked as $[0/1]$ are the $\epsilon=1$ values of the Pad\'e approximants $\Delta_C=a_0/(1-\epsilon\,a_1/a_0)$. 

\begin{table}[h,t,b]
\centering
\caption{Estimated ${d=3}$~values of $\Delta_C/n$. In the easy-axes cases $(m_{c,c},m_{\mathrm{D},\mathrm{D}},n)=(1,n-1,n)$ and $(0,n-1,n)$, the numbers of modes with Dirichlet boundary conditions only on one boundary plane satisfy $m_{c,\mathrm{D}}=m_{\mathrm{D},c}=0$ and $m_{c,\mathrm{D}}+m_{\mathrm{D},c}=1$, respectively. In the cases $(n,m_{c,c},m_{\mathrm{D},\mathrm{D}})=(2,0,0)$  and $(3,0,1)$ it is understood that $m_{c,\mathrm{D}}=m_{\mathrm{D},c}=1$. The values in the column labeled MC refer to Monte Carlo results reported in Refs.~\cite{Huc07}, \cite{VGMD07}, \cite{NI85}, and \cite{INW86} (which agree with each other).} \label{tab:Deltaest}
\medskip
\begin{tabular}{ccccccccc}\hline
$n$&$m_{c,c}$&$m_{\mathrm{D},\mathrm{D}}$&$O(\epsilon^0)$&$O(\epsilon)$&$O(\epsilon^{3/2})$&$[0/1]$&MC\\\hline
1&1&0&$-0.00685$&$-0.01166$&$-0.02243$&$-0.02293$&--\\
1&0&1&$-0.00685$&$-0.01166$&&$-0.02293$&$-0.015$\\
1&0&0&$+0.00600$&$+0.01282$&&$-0.04352$&\\
2&1&1&$-0.00685$&$-0.01109$&$-0.01817$&$-0.01793$&\\
2&0&2&$-0.00685$&$-0.01109$&&$-0.01793$&$-0.015(6)$\\
2&0&1&$-0.00043$&$-0.00003$&&$-0.00022$&\\
2&0&0&$+0.00600$&$+0.01296$&&$-0.03711$&\\
3&1&2&$-0.00685$&$-0.01062$&$-0.01634$&$-0.01522$&\\
3&0&2&$-0.00257$&$-0.00389 $&$$&$-0.00528$&\\
3&0&1&$+0.00171$&$+0.00401$&&$-0.00502$&\\
$n\to\infty$&1&$n-1$&$-0.00685$&$-0.00595$&$-0.00595$&$-0.00605$\\
$n\to\infty$&0&$n-1$&$-0.00685$&$-0.00595$&&$-0.00605$\\
$n\to\infty$&0&$n$&$-0.00685$&$-0.00595$&&$-0.00605$\\
\hline
\end{tabular}
\end{table}

In cases with $m_{c,c}=0$ (where no zero modes are present in Landau theory at $T_{c,\infty}$), we expect the quality of the extrapolations to be comparable with the original $O(\epsilon)$ estimates of Krech and Dietrich \cite{KD91,KD92a} for the isotropic $n>1$ cases with $m_{\mathrm{D},\mathrm{D}}=n$ and $m_{c,\mathrm{D}}=n$, respectively. In the Ising case with $n=m_{\mathrm{D},\mathrm{D}}=1$ they used a Pad\'e $[1/1]$ approximant whose additional parameter was fixed by the requirement that the exact value $\Delta_C(d=2)=-\pi/48\simeq-0.65$ be reproduced in addition to the $\epsilon$~expansion to first order. This gave the estimate $\Delta_C(n=m_{\mathrm{D},\mathrm{D}}=1,d=3)\simeq -0.015$, which is in very good agreement with Monte Carlo results. We have refrained from giving this value in Table~\ref{tab:Deltaest}  in order to have a better comparison with the $n>1$ cases, for which analogous exact results that could be used for improved Pad\'e  approximants are not available to our knowledge. 

Obtaining reliable extrapolations in the cases with $m_{c,c}>0$, where  fractional powers of $\epsilon$ (modulo logarithms) appear in the small-$\epsilon$ expansions, is an even greater challenge. To this end a better understanding of these series would be needed. Therefore, reliable information about Casimir amplitudes from alternative sources such as Monte Carlo simulations, exact model calculations, and other approximate RG calculations would be very useful.

We also included estimates for $\Delta_C/n$ in the large-$n$ limit. This limit is known to be equivalent to an appropriate spherical model with $z$-dependent constraints \cite{Kno73}. Its exact solution leads to a self-consistent Euclidean Schr\"odinger equation \cite{BM77} whose analytical solution is not known for finite $L$, not even at $T_{c,\infty}$. Recent numerical work~\cite{Comtesse09} gave the result $\lim_{n\to\infty}\Delta_C(d=3)/n\simeq -0.012(1)$ for open boundary conditions, corresponding to the case $m_{\mathrm{D},\mathrm{D}}=n$. According to previous work \cite{DGS06,GD08,DC09}, the uncertainties of naive extrapolations of the small-$\epsilon$ expansions become significant as $n$ increases. That our extrapolation  gives a value at $d=3$ which is roughly $50\%$ of this numerical large-$n$ result is therefore not unexpected. On the other hand, we expect our small-$\epsilon$ expansion results to be in conformity with the large-$n$ limits in $d=4-\epsilon$ dimensions since we were able to show this for our previously published analogous  results for periodic boundary conditions \cite{DGS06,GD08}.

\section{Summary and concluding remarks}
 \label{sec:concl}

In this work we studied effective interactions induced by thermal fluctuations in a medium that is confined by two parallel $(d-1)$-dimensional boundary planes at its $d$-dimensional bulk critical point. These fluctuation-induced interactions manifest themselves through a contribution to the free energy per film area $A\to \infty$ that depends on the separation $L$ of the confining planes. At the critical temperature $T_{c,\infty}$ of the $L=\infty$ (bulk) system, the residual free energy densities $f_{\mathrm{res}}(L;T)$ introduced in equation~\eqref{eq:Fsplit} exhibit the asymptotic power-law decays \eqref{eq:frescrit} in $L$. The associated fluctuation-induced (``Casimir'') forces per area,
\begin{equation}
   \label{eq:FC}
\mathcal{F}_{C}(L;T)=-k_B T\,\frac{\partial f_{\mathrm{res}}(L;T)}{\partial L},
\end{equation}
therefore become long-ranged at $T_{c,\infty}$  \cite{Kre94,KD91,KD92a,FdG78}. They decay as $(d-1)\Delta_C^{(\wp)}L^{-d}$, and hence are attractive or repulsive depending on whether the amplitudes $\Delta_C^{(\wp)}$ are negative or positive.

These forces $\mathcal{F}_{C}$ are the analogs of the familiar Casimir forces in quantum electrodynamics (QED) between two ideally conducting grounded metallic parallel plates at separation $L$ caused by  vacuum fluctuations of the electromagnetic field \cite{Cas48,BMM01}. Just as the latter forces, their thermal analogs $\mathcal{F}_C$ depend on gross features of the medium (such as space dimension $d$, number of components $n$ of the order parameter, etc) and of the confining plates (such as boundary conditions). In the QED case, the interaction of the electrodynamic field with matter (i.e., the metallic plates) usually  is taken into account only through the choice of appropriate boundary conditions. Hence the problem reduces to the study of free field theories in bounded geometries under given  boundary conditions $\wp$. Common physically relevant choices are the combinations (D,D), (N,N), (D,N), and (N,D) of Dirichlet (D) and Neumann (N) boundary conditions.  Massless Gaussian field theories have the simplifying feature that when these boundary conditions 
are imposed  (at the mesoscopic scale at which the continuum description applies), they remain valid at larger length scales. The reason is that Dirichlet and Neumann boundary conditions at either one of the boundary planes correspond to infrared stable and unstable fixed points of the corresponding boundary field theories, respectively \cite{Die86a,Die97}. Choosing for each component $\phi_\alpha$ one of these boundary conditions at both planes $\mathfrak{B}_j$ fixes the Casimir amplitude of the Gaussian theory for given $d$ and $n$ uniquely; the corresponding values $\Delta_{C,\mathrm{Gauss}}$ are given in equation~\eqref{eq:DeltaGauss}. 

On the other hand, if instead of a Neumann or Dirichlet boundary condition a Robin boundary condition $\partial_n\phi_\alpha=\mathring{c}_\alpha^{(j)}\phi_\alpha$ with $0<\mathring{c}_\alpha^{(j)}<\infty$ is imposed at $\mathfrak{B}_j$, then this mesoscopic boundary condition does not remain valid on larger scales but  turns into an asymptotic Dirichlet boundary condition in the large-length-scale  limit \cite{SD08,Die86a,Die97}. This means that effective scale-dependent Casimir amplitudes,  depending on the scaled argument $\mathring{c}_\alpha^{(j)}L$, result whenever such a Robin boundary condition is involved. 

The study of critical Casimir forces below the upper critical dimension $d^*=4$ requires going beyond the Landau and Gaussian approximations. Then important changes of the scenario just described occur. They result from the fact that the $\phi^4$ interaction shifts the Gaussian fixed point with $\mathring{c}_\alpha^{(j)}=0$ (all $\alpha$), at which a Neumann boundary condition applies  for $\bm{\phi}$ at $\mathfrak{B}_j$ on the level of this free theory, to a nonuniversal (cutoff dependent) value $\mathring{c}_{\mathrm{sp}}$ [see equation~\eqref{eq:srep}]. The associated shifted fixed points are the ones of the $O(n)$ invariant theory called ``special''. As already recalled in the Introduction, their physical significance is to describe the so-called isotropic special surface transitions \cite{Die86a,Die97}, which occur (in sufficiently high space dimensions) for critically enhanced surface interactions. 

Accordingly, Neumann boundary conditions do no longer correspond to RG fixed points, neither to stable nor unstable ones. The consequences  are threefold: (i) Neumann boundary conditions that hold on a mesoscopic scale do not persist on longer length scales. The asymptotic boundary conditions on large length scales are determined by the RG fixed points to whose basins of attraction the values of the surface enhancement variables $c_\alpha^{(j)}$ belong. For simple Ising models with nearest-neighbor interactions whose bonds have different strengths in the two boundary layers and elsewhere, Neumann boundary conditions are expected to lie in the basins of attraction of the fixed points with $c_\alpha^{(j)}=\infty$ at which asymptotic Dirichlet boundary conditions prevail \cite{SD08}. (ii) Since the analogs of the Gaussian fixed points with $\mathring{c}^{(j)}_\alpha=0$ ---- the special fixed points --- are located in the space of bare interaction constants at a nonuniversal value $\mathring{c}_{\mathrm{sp}}$, the mesoscopic boundary condition $\partial_n\bm{\phi}=\mathring{c}_{\mathrm{sp}}\bm{\phi}$ one has for critically enhanced surface interactions is also nonuniversal. (iii) Again, this boundary condition does not persist on longer length scales. On large length scales $\ll L, \xi_\infty$, a characteristic algebraic near-boundary behavior holds in this case of critical enhancement, as predicted by the boundary operator expansion (cf.\ section~\ref{sec:isoren}).

The main purpose of this paper was to generalize previous investigations of critical Casimir amplitudes for systems with $n$-component order parameters in a film geometry by allowing for  $O(n)$-symmetry breaking quadratic boundary contributions in the Hamiltonian. Aside from the above-mentioned Gaussian results~\eqref{eq:DeltaGauss}, our main findings are the small-$\epsilon$ series expansions of the Casimir amplitudes $\Delta_C$ gathered in equations~\eqref{eq:DeltafresR}--\eqref{eq:a3ov2} and the $d=3$ estimates presented in Table~\ref{tab:Deltaest}. There are several good reasons for the generalization we made.

First, it is not uncommon that surface spin anisotropies occur in systems whose bulk critical behavior belongs to the universality class of the $O(n)$ model. For this reason and from a fundamental point of view, there is therefore interest in a study of fluctuation-induced interactions in such systems. Second, even in the absence of such symmetry breaking boundary terms, one can consider boundary conditions on the mesoscopic scale of the continuum field theory that break the $O(n)$ symmetry. Common ways of breaking the internal $\mathbb{Z}_2$ symmetry in the scalar ($n=1$) case are choices of $\pm \pm$ and $\pm\mp$ boundary conditions, for which the values of the order parameter at the boundary planes $\mathfrak{B}_j$ are fixed to nonvanishing values of  same or opposite signs. Such boundary conditions are of direct relevance for fluctuation-induced forces that are mediated by binary liquid mixtures near their consolute point \cite{Kre97,VGMD07,FdG78,ML00,FYP05,RBM07,HHGDB08}. In the continuous symmetry case $n>1$, it is natural to consider also twisted boundary conditions for which the local order parameters $\bm{\phi}|_{\mathfrak{B}_j}$ at the two boundary planes are aligned along different directions but have the same fixed magnitude \cite{Bar83,FBJ73,CL95}. Rather than by such inhomogeneous boundary conditions, one can also break the symmetry by homogeneous boundary conditions. Thus one can choose different Robin boundary conditions for the $n$ components of the order parameter. This is precisely the situation we were concerned with in this paper. Just as in the absence of $O(n)$-symmetry breaking boundary terms,  the boundary conditions that hold on a given mesoscopic scale do not necessarily hold on larger scales, let alone asymptotically in the large-length-scale limit \cite{SD08,Die86a,Die97}. Determination of the boundary conditions in the large-length-scale limit requires the understanding of the respective fixed points of the associated boundary field theory and the boundary operator expansions. These issues were dealt with in some detail in Sections~\ref{sec:isoren} and \ref{sec:msp}.

An important difference between symmetry breaking by the mentioned inhomogeneous ($\pm\pm$, $\pm\mp$
or twisted) boundary conditions and the latter homogeneous ones should be realized. The former, which can be realized by symmetry breaking linear boundary terms in the Hamiltonian --- i.e., surface fields $\bm{h}_j =h_j\hat{\bm{h}}_j$ whose components $h_j$ along certain directions specified by the unit vectors $\hat{h}_j$ tend to infinity --- imply a \emph{nonzero order-parameter profile} $\bm{m}(z)=\langle\bm{\phi}(\bm{y},z)\rangle $ \emph{both above and below} the bulk critical temperature $T_{c,\infty}$ even in Landau theory. In the case of quadratic $O(n)$-symmetry breaking boundary terms (or corresponding homogeneous boundary conditions), the order parameter profile vanishes above $T>T_{c,\infty}$, unless some enhancement variables  are supercritically enhanced so that in a temperature region above $T_{c,\infty}$ spontaneous symmetry breaking occurs and a nonzero profile $\bm{m}(z)$ results. When this possibility is ruled out because all surface interactions are subcritically or at most critically enhanced, the Casimir force that results for temperatures $T\ge T_{c,\infty}$ is \emph{entirely due to fluctuations} and vanishes in Landau theory. In the case of linear symmetry-breaking surface terms (or corresponding inhomogeneous boundary conditions), these forces generically vanish neither above nor below nor at $T_{c,\infty}$.  Thus, at and above $T_{c,\infty}$ they are \emph{not} purely fluctuation induced, a property they share with the Casimir force below $T_{c,\infty}$ in the case of homogeneous boundary conditions. One may question whether the term ``Casimir force'' is appropriate when a force is not entirely due to fluctuations (as the original QED Casimir force is), but it has become customary to use it even in such cases \cite{HHGDB08}. The critical Casimir interactions we studied above are all completely fluctuation induced and hence  more compelling analogs of those known from QED. This provides  a further reason for  investigating them.

A fourth reason is that  one does not expect special surface transitions to be possible in semi-infinite geometry for systems with $O(n)$ symmetric Hamiltonians and short-ranged interactions. The introduction of quadratic symmetry-breaking boundary terms corresponding to easy-axes spin anisotropies opens up the possibility of having large-length-scale boundary conditions that are governed by anisotropic special fixed points. A fifth reason is that the cases with $m_{c,c}>0$ (involving zero modes) provide further examples of the breakdown of the $\epsilon$ expansion for critical Casimir amplitudes.

A natural question to ask is how the Casimir interactions investigated in this work can be checked by experiments and simulations. Previous experimental work on thermodynamic Casimir interactions has focused on fluid systems \cite{ML00,FYP05,RBM07,HHGDB08,GC99}. The advantage of fluid systems is that the number of degrees of  freedom of the medium (fluid) between macroscopic objects (two walls, say) can vary; one has realizations of grand canonical ensembles. Obvious candidates to which the $O(n)$ model with symmetry breaking quadratic boundary terms studied in this work might apply are magnetic systems. Unlike fluids, magnetic systems do not lend themselves to direct or indirect types of measurements of thermodynamic forces of the kind used for fluid systems. Nevertheless,  finite-size and residual free energies (as well as their temperature derivatives) are observable quantities that in principle can be measured. As far as Monte Carlo calculations are concerned, we do not see qualitatively new challenges. The simulation techniques that are currently in use for  studying lattice $n$-vector models should also work when easy-axes spin anisotropies in the boundary layers are included. Of course, in cases where some surface bonds are critically enhanced, careful --- and cumbersome --- determinations of the corresponding critical enhancements are required. We hope that the present work will stimulate such Monte Carlo investigations, as well as experiments.

\ack
We greatfully acknowledge partial support by the Deutsche Forschungsgemeinschaft under Grant No.\ Di-378/5.
\appendix
\section{Calculation of required integrals}\label{app:requint}

In this Appendix we compute the integrals $I_{1}^{(A,B)}(L;\mathring{\tau}=0)$
and $I_{2}^{(\mathrm{A}_\alpha,\mathrm{B}_{\alpha};\mathrm{A}_\beta,\mathrm{B}_{\beta})}(L;\mathring{\tau}=0)$ defined
by Eqs.~\eqref{eq:I1def} and \eqref{eq:Iwp2}, respectively.

To compute $I_{1}^{(\mathrm{A},\mathrm{B})}(L;0)$, note that the spectral decomposition of the free propagator $G_{L,\psi,\alpha}(\bm{p};z,z)$ is given by Eq.~\eqref {eq:Gfreespecrep}, except that we must leave out the $r=0$ summand with eigenvalue $\hat{k}_0^2=0$ when $(\mathrm{A}_\alpha,\mathrm{B}_\alpha)=(\mathrm{N},\mathrm{N})$. Substituting this expression into Eq.~\eqref{eq:I1def}, we can perform the integration over $\bm{p}$ and exploit the fact that the eigenfunctions $\varphi_r^{(A,B)}(\zeta)$ given in Eq.~\eqref{eq:varphiAB} are orthonormal on the interval $[0,1]$ to obtain
\begin{equation}
\label{eq:I1calc}
I_{1}^{(\mathrm{A},\mathrm{B})}(L;0)=-\frac{2^{1-d}\pi^{(3-d)/2}L^{3-d}}{\Gamma[(d-1)/2]\cos(d\pi/2)}
\sideset{}{'}\sum_r\hat{k}_r^{d-3}
\end{equation}
where the prime on $\sum_r'$ indicates omission of the $r=0$ summand when $(\mathrm{A}_\alpha,\mathrm{B}_\alpha)=(\mathrm{N},\mathrm{N})$. 

Since the eigenvalues $\hat{k}_r^2$ depend on the boundary conditions $(\mathrm{A}_\alpha,\mathrm{B}_\alpha)$, so do the series $\sum_r'$. One finds 
\begin{equation}
\sideset{}{'}\sum_r\hat{k}_{r}^{ d-3}=\frac{\zeta(3-d)}{\pi^{3-d}}\times 
\begin{cases}
1 & \text{for }  (\mathrm{A}_\alpha,\mathrm{B}_\alpha)=(\mathrm{N},\mathrm{N})\,,\\
1 & \text{for } (\mathrm{A}_\alpha,\mathrm{B}_\alpha)=(\mathrm{D},\mathrm{D})\,,\\
2^{3-d}-1 & \text{for } (\mathrm{A}_\alpha,\mathrm{B}_\alpha)=(\mathrm{D},\mathrm{N})\text{ or } (\mathrm{N},\mathrm{D})\,,
\end{cases}
\end{equation}
which together with Eq.~\eqref{eq:I1calc} yields the results stated in Eq.~\eqref{eq:I1res}.

An analogous calculation yields
\begin{equation}
  \label{eq:I2app}
I_{2}^{(\mathrm{A}_\alpha,\mathrm{B}_{\alpha};\mathrm{A}_\beta,\mathrm{B}_{\beta})}(L;0)  =  \frac{2^{2-2d}\pi^{3-d}L^{6-2d}}{\Gamma^{2}[(d-1)/2]\cos^{2}(d\pi/2)}\sideset{}{'}\sum_{r_1}\sideset{}{'}\sum_{r_2}\hat{k}_{r_1}^{d-3}\hat{k}_{r_2}^{d-3}\,,
\end{equation}
where we used the fact that
\begin{eqnarray}
\lefteqn{\int_{0}^{1}\mathrm{d}{\zeta}\big[\varphi_{r_1}^{(\mathrm{A}_\alpha,\mathrm{B}_\alpha)}(\zeta)\big]^2\,\big[\varphi_{r_2}^{(\mathrm{A}_\beta,\mathrm{B}_\beta)}(\zeta) \big]^2}&&\nonumber\\  &=&
\begin{cases}
1+\frac{1}{2}\,\delta_{r_1r_2}&\text{for } \alpha=\beta,\\
1&\text{for }(\mathrm{A}_\beta,\mathrm{B}_\beta)\ne(\mathrm{A}_\alpha,\mathrm{B}_\alpha)=(\mathrm{D},\mathrm{N}), (\mathrm{N},\mathrm{D}),\\
1-\frac{1}{2}\,\delta_{r_1r_2}&\text{for }\mathrm{A}_\alpha=\mathrm{B}_{\alpha}=\mathrm{N} \text{ and }\mathrm{A}_\beta=\mathrm{B}_{\beta}=\mathrm{D}.
\end{cases}
\end{eqnarray}

The summations $\sum'_{r_1}$ and $\sum'_{r_2}$ can be carried out in a straightforward fashion. As results one obtains
\begin{equation}
I_{2}^{(\mathrm{A,A};\mathrm{B,B})}(L;0)=\frac{L^{4-2d}2^{1-2d}\pi^{d-3}\big[2\zeta^{2}(3-d)+(2\delta_{\mathrm{A}\mathrm{B}}-1)\zeta(6-2d)\big]}{\Gamma^{2}[(d-1)/2]\cos^{2}(d\pi/2)},
\end{equation}
\begin{eqnarray}
I_{2}^{(\mathrm{D,N};\mathrm{D,N})}(L;0)&
=&\frac{L^{4-2d}2^{1-2d}\pi^{d-3}\big[2(2^{3-d}-1)^{2}\zeta^{2}(3-d)+(2^{6-2d}-1)\zeta(6-2d)\big]}{\Gamma^{2}[(d-1)/2]\cos^{2}(d\pi/2)},\nonumber\\
\end{eqnarray}
and
\begin{equation}
I_{2}^{(\mathrm{D,N};\mathrm{N,N})}(L;0)=I_{2}^{(\mathrm{D,N};\mathrm{D,D})}(L;0)=\frac{L^{4-2d}2^{2-2d}\pi^{d-3}(2^{3-d}-1)\zeta^{2}(3-d)}{\Gamma^{2}[(d-1)/2]\cos^{2}(d\pi/2)}.
\end{equation}

We need the $\epsilon$ expansions of these integrals only to zeroth order. They read
\begin{equation}
I_{2}^{(\mathrm{A,A};\mathrm{B,B})}(L;0)=\frac{1}{2304L^{4}}+O(\epsilon),\;\;\mathrm{A},\mathrm{B}=\mathrm{N},\mathrm{D},
\end{equation}
\begin{equation}
I_{2}^{(\mathrm{D,N};\mathrm{D,N})}(L;0)=\frac{1}{9216L^{4}}+O(\epsilon),
\end{equation}
and
\begin{equation}
I_{2}^{(\mathrm{D,N};\mathrm{N,N})}(L;0)=I_{2}^{(\mathrm{D,N};\mathrm{D,D})}(L;0)=-\frac{1}{4608L^{4}}+O(\epsilon).
\end{equation}

\section{Derivation of the representation~\eqref{eq:GAB} of $\bm{G_{L}^{(\mathrm{D,N})}(\bm{x}_{1};\bm{x}_{2})}$}\label{app:GLDsp}

Here we give a derivation of Eq.~\eqref{eq:GAB} based on Poisson's summation formula~\eqref{eq:PoissonSumFormula}. We start from the spectral representation~\eqref {eq:Gfreespecrep}, use this summation formula, and shift the integration variable $k$ by $\pi/2L$. We thus arrive at
 \begin{equation}
G_{L,\psi}^{(\mathrm{D,N})}(\bm{x}_{1};\bm{x}_{2})=\frac{1}{\pi}
\sum_{j=-\infty}^{\infty}(-1)^{j}\int_{\bm{p}}^{(d-1)}\int_{-\infty}^{\infty}\mathrm{d}{k}\,\frac{\sin(kz_{1})\sin(kz_{2})}{p^{2}+k^{2}+\mathring{\tau}}\,
\mathrm{e}^{2\mathrm{i} jLk+\mathrm{i}\bm{p}\cdot\bm{y}_{12}}.
\end{equation}
The $k$~integration can be performed. This gives
 \begin{eqnarray}
G_{L}^{(\mathrm{D,sp})}(\bm{x}_{1};\bm{x}_{2})
  &=&  \sum_{j=-\infty}^{\infty}(-1)^{j}\int_{\bm{p}}^{(d-1)}\frac{\exp(\mathrm{i}\bm{p}\cdot\bm{y}_{12})}{2\sqrt{p^{2}+\mathring{\tau}}}\Big[\mathrm{e}^{-|2Lj+z_{2}-z_{1}|\sqrt{p^{2}+\mathring{\tau}}}\nonumber\\ &&
  \phantom{\sum_{j=-\infty}^{\infty}(-1)^{j}\int_{\bm{p}}^{(d-1)}\frac{1}{2\sqrt{p^{2}+\mathring{\tau}}}}
 \strut-\mathrm{e}^{-|2Lj+z_{2}+z_{1}|\sqrt{p^{2}+\mathring{\tau}}}\Big].
\end{eqnarray}
 When the result is expressed in terms of the free massive bulk propagator
\begin{equation}
G_{\infty}^{(d)}(\bm{x}|\mathring{\tau})=\int_{\bm{p}}^{(d-1)}\frac{1}{2\sqrt{p^{2}
+\mathring{\tau}}}\,\mathrm{e}^{-|z|\sqrt{p^{2}+\mathring{\tau}}}\,\mathrm{e}^{\mathrm{i}\bm{p}\cdot\bm{y}},
\end{equation}
the representation~\eqref{eq:GAB} of $G_{L,\psi}^{(\mathrm{D,N})}(\bm{x}_{1};\bm{x}_{2})$ follows.


\begin{thebibliography}{10}
\expandafter\ifx\csname url\endcsname\relax
  \def\url#1{\texttt{#1}}\fi
\expandafter\ifx\csname urlprefix\endcsname\relax\def\urlprefix{URL }\fi

\bibitem{Fis71}
M.~E. Fisher, The theory of critical point singularities, in: M.~S. Green
  (Ed.), Critical Phenomena, Proceedings of the 51st.\ Enrico Summer School,
  Varenna, Italy, Academic, London, 1971, pp. 73--98.

\bibitem{Bar83}
M.~N. Barber, Finite-size scaling, in: C.~Domb, J.~L. Lebowitz (Eds.), Phase
  Transitions and Critical Phenomena, Vol.~8, Academic, London, 1983, pp.
  145--266.

\bibitem{Bin87}
K.~Binder, Finite size effects on phase transitions, Ferroelectrics 73 (1987)
  43--67.

\bibitem{Pri90}
V.~Privman, Finite-size scaling theory, in: V.~Privman (Ed.), Finite Size
  Scaling and Numerical Simulation of Statistical Systems, World Scientific,
  Singapore, 1990, Ch.~1.

\bibitem{BDT00}
J.~G. Brankov, D.~M. Dantchev, N.~S. Tonchev, Theory of Critical Phenomena in
  Finite-Size Systems --- Scaling and Quantum Effects, World Scientific,
  Singapore, 2000.

\bibitem{Kre94}
M.~Krech, {C}asimir Effect in Critical Systems, World Scientific, Singapore,
  1994.

\bibitem{Doh08}
V.~Dohm, Diversity of critical behavior within a universality class, Phys. Rev.
  E 77~(6).

\bibitem{DC09}
H.~W. Diehl, H.~Chamati, Dynamic critical behavior of model a in films:
  Zero-mode boundary conditions and expansion near four dimensions, Phys. Rev.
  B 79~(10) (2009) 104301, arXiv:0810.5244.
\newline\urlprefix\url{http://link.aps.org/abstract/PRB/v79/e104301}

\bibitem{KD91}
M.~Krech, S.~Dietrich, Finite-size scaling for critical films, Phys. Rev. Lett.
  66 (1991) 345--348, [Erratum {\bf 67}, 1055 (1991)].

\bibitem{KD92a}
M.~Krech, S.~Dietrich, Free energy and specific heat of critical films and
  surfaces, Phys. Rev. A 46~(4) (1992) 1886--1922.

\bibitem{KD92b}
M.~Krech, S.~Dietrich, Specific heat of critical films, the {C}asimir force,
  and wetting films near critical end points, Phys. Rev. A 46~(4) (1992)
  1922--1941.

\bibitem{SD08}
F.~M. Schmidt, H.~W. Diehl, Crossover from attractive to repulsive {C}asimir
  forces and vice versa, Phys. Rev. Lett. 101~(10) (2008) 100601.
\newline\urlprefix\url{http://link.aps.org/abstract/PRL/v101/e100601}

\bibitem{NI85}
M.~P. Nightingale, J.~O. Indekeu, Effect of criticality on wetting layers,
  Phys. Rev. Lett. 54~(1) (1985) 1824--1827.

\bibitem{INW86}
J.~O. Indekeu, M.~P. Nightingale, W.~V. Wang, Finite-size interaction
  amplitudes and their universality: Exact, mean-field, and
  renormalization-group results, Phys. Rev. B 34~(1) (1986) 330--342.

\bibitem{KL96}
M.~Krech, D.~P. Landau, {C}asimir effect in critical systems: A {M}onte {C}arlo
  simulation, Phys.\ Rev.\ E 53 (1996) 4414--4423.

\bibitem{Kre97}
M.~Krech, {C}asimir forces in binary liquid mixtures, Phys. Rev. E 56~(2)
  (1997) 1642--1659.

\bibitem{DK04}
D.~Dantchev, M.~Krech, The critical {C}asimir force and its fluctuations in
  lattice spin models: exact and {M}onte {C}arlo results, Phys. Rev. E 69~(4)
  (2004) 046119--1--20, cond-mat 0402238.

\bibitem{DDG06}
D.~Dantchev, H.~W. Diehl, D.~Gr{\"u}neberg, Excess free energy and {C}asimir
  forces in systems with long-range interactions of van der {W}aals type:
  General considerations and exact spherical-model results, Phys. Rev. E 73
  (2006) 016131--1--26, cond-mat/0510405.

\bibitem{Huc07}
A.~Hucht, Thermodynamic {C}asimir effect in $^4${He} films near ${T}_\lambda$:
  {M}onte {C}arlo results, Phys. Rev. Lett. 99~(18) (2007) 185301.
\newline\urlprefix\url{http://link.aps.org/abstract/PRL/v99/e185301}

\bibitem{VGMD07}
O.~Vasilyev, A.~Gambassi, A.~Macio{\l}ek, S.~Dietrich, {M}onte {C}arlo
  simulation results for critical {C}asimir forces, Europhys. Lett. 80~(6)
  (2007) 60009 (6pp).
\newline\urlprefix\url{http://stacks.iop.org/0295-5075/80/60009}

\bibitem{Bin83}
K.~Binder, Critical behaviour at surfaces, in: C.~Domb, J.~L. Lebowitz (Eds.),
  Phase Transitions and Critical Phenomena, Vol.~8, Academic, London, 1983, pp.
  1--144.

\bibitem{Die86a}
H.~W. Diehl, Field--theoretical approach to critical behaviour at surfaces, in:
  C.~Domb, J.~L. Lebowitz (Eds.), Phase Transitions and Critical Phenomena,
  Vol.~10, Academic, London, 1986, pp. 75--267.

\bibitem{Die97}
H.~W. Diehl, The theory of boundary critical phenomena, Int.\ J.\ Mod.\ Phys.\
  B 11 (1997) 3503--3523, cond-mat/9610143.

\bibitem{BD94}
T.~W. Burkhardt, H.~W. Diehl, Ordinary, Extraordinary, and normal surface
  transitions: Extraordinary--normal equivalence and simple explanation of $
  |{T}-{T}_c|^{2-\alpha}$ singularities, Phys.\ Rev.\ B 50~(6) (1994)
  3894--3898.

\bibitem{Die94a}
H.~W. Diehl, Critical adsorption of fluids and the equivalence of extraordinary
  and normal surface transitions, Ber.\ Bunsenges.\ Phys.\ Chem. 98 (1994)
  466--471.

\bibitem{DE82}
H.~W. Diehl, E.~Eisenriegler, Irrelevance of surface anisotropies for critical
  behavior near free surfaces, Phys.\ Rev.\ Lett. 48 (1982) 1767--1768.

\bibitem{DE84}
H.~W. Diehl, E.~Eisenriegler, Effects of surface exchange anisotropies on
  magnetic critical and multicritical behavior at surfaces, Phys.\ Rev.\ B 30
  (1984) 300--314.

\bibitem{DGS06}
H.~W. Diehl, D.~Gr{\"u}neberg, M.~A. Shpot, Fluctuation-induced forces in
  periodic slabs: Breakdown of $\epsilon$ expansion at the bulk critical point
  and revised field theory, Europhys. Lett. 75 (2006) 241--247,
  cond-mat/0605293.

\bibitem{GD08}
D.~Gr{\"u}neberg, H.~W. Diehl, Thermodynamic {C}asimir effects involving
  interacting field theories with zero modes, Phys. Rev. B 77~(11) (2008)
  115409, {arXiv:0710.4436}.
\newline\urlprefix\url{http://link.aps.org/abstract/PRB/v77/e115409}

\bibitem{RS02}
A.~Romeo, A.~A. Saharian, {C}asimir effect for scalar fields under {R}obin
  boundary conditions on plates, J. Phys. A 35 (2002) 1297--1320.

\bibitem{Sch08}
F.~M. Schmidt, {Kritischer Casimir Effekt bei Robin-Randbedingungen}, Diploma
  thesis, Fachbereich Physik, Universit\"at Duisburg-Essen, Duisburg (February
  2008).

\bibitem{MF53}
P.~M. Morse, H.~Feshbach, Methods of Theoretical Physics, Part I, McGraw-Hill,
  New York, 1953.

\bibitem{DD80}
H.~W. Diehl, S.~Dietrich, Scaling laws and surface exponents from
  renormalization--group equations, Phys.\ Lett. 80A (1980) 408--412.

\bibitem{Sym81}
K.~Symanzik, Schr{\"o}dinger representation and {C}asimir effect in
  renormalizable quantum field theory, Nucl.\ Phys.\ B 190 (1981) 1--44.

\bibitem{DD81a}
H.~W. Diehl, S.~Dietrich, Field--theoretical approach to static critical
  phenomena in semi--infinite systems, Z.\ Phys.\ B: Condens. Matter 42 (1981)
  65--86, erratum: {\bf 43}, 281 (1981).

\bibitem{DD81b}
H.~W. Diehl, S.~Dietrich, Field--theoretical approach to multicritical behavior
  near free surfaces, Phys.\ Rev.\ B 24 (1981) 2878--2880.

\bibitem{DD83a}
H.~W. Diehl, S.~Dietrich, Multicritical behaviour at surfaces, Z.\ Phys.\ B:
  Condens. Matter 50 (1983) 117--129.

\bibitem{DE85}
H.~W. Diehl, E.~Eisenriegler, Effects of surface exchange anisotropies on
  magnetic critical and multicritical behavior at surfaces, in: L.~Garrido
  (Ed.), Applications of Field Theory to Statistical Mechanics, Vol. 216 of
  Lecture Notes in Physics, Proceedings of the VIII Sitges Conference,
  Springer-Verlag, Berlin, 1985, pp. 343--347.

\bibitem{DBN05}
Y.~Deng, H.~W.~J. Bl{\"o}te, M.~P. Nightingale, Surface and bulk transitions in
  three-dimensional $O(n)$ models, Physical Review E (Statistical, Nonlinear, and
  Soft Matter Physics) 72~(1) (2005) 016128.
\newline\urlprefix\url{http://link.aps.org/abstract/PRE/v72/e016128}

\bibitem{Kre00}
M.~Krech, Surface scaling behavior of isotropic Heisenberg systems: Critical
  exponents, structure factor, and profiles, Phys. Rev. B 62~(10) (2000)
  6360--6371.

\bibitem{DKK02}
H.~W. Diehl, M.~Krech, H.~Karl, Dynamic surface critical behavior of isotropic
  {H}eisenberg ferromagnets: Boundary conditions, renormalized field theory,
  and computer simulation results, Phys. Rev. B 66 (2002) 024408,
  cond-mat/0203368.

\bibitem{BZJ85}
E.~Br{\'e}zin, J.~Zinn-Justin, Finite size effects in phase transitions, Nucl.\
  Phys.\ B 257 (1985) 867--893.

\bibitem{RGJ85}
J.~Rudnick, H.~Guo, D.~Jasnow, Finite-size scaling and the renormalization
  group, J. Stat. Phys. 41 (1985) 353--373.

\bibitem{Kno73}
H.~J.~F. Knops, Infinite spin dimensionality limit for nontranslationally
  invariant interactions, J. Math. Phys. 14~(12) (1973) 1918--1920.

\bibitem{BM77}
A.~J. Bray, M.~A. Moore, Critical behaviour of semi--infinite systems, J.\
  Phys.\ A 10 (1977) 1927--1961.

\bibitem{Comtesse09}
D.~Comtesse, A.~Hucht, D.~Gr\"uneberg, Thermodynamic {C}asimir effect in the
  large-$n$ limit, arXiv:0904.3661v1.

\bibitem{FdG78}
M.~E. Fisher, P.-G. de~Gennes, Ph{\'e}nom{\`e}nes aux parois dans un
  m{\'e}lange binaire critique, C.\ R.\ S{\'e}ances.\ Acad.\ Sci.\ S{\'e}rie B
  287 (1978) 207--209.

\bibitem{Cas48}
H.~B.~G. Casimir, On the attraction between two perfectly conducting plates,
  Proc. K. Ned. Akad. Wet. B51 (1948) 793--795.

\bibitem{BMM01}
M.~Bordag, U.~Mohideen, V.~M. Mostepanenko, New developments in the {C}asimir
  effect, Phys. Rep. 353 (2001) 1--205.
  
\bibitem{ML00}
A.~Mukhopadhyay, B.~M. Law, {C}asimir effect in critical films of binary liquid
  mixtures, Phys. Rev. E 62 (2000) 5201.

\bibitem{FYP05}
M.~Fukuto, Y.~F. Yano, P.~S. Pershan, Critical {C}asimir effect in
  three-dimensional {I}sing systems: measurements on binary wetting films,
  Phys. Rev. Lett. 94 (2005) 135702--1--4.

\bibitem{RBM07}
S.~Rafai, D.~Bonn, J.~Meunier, Repulsive and attractive critical casimir
  forces, Physica A 386~(1) (2007) 31--35.

\bibitem{HHGDB08}
C.~Hertlein, L.~Helden, A.~Gambassi, S.~Dietrich, C.~Bechinger, Direct
  measurement of critical {C}asimir forces, Nature 451 (2008) 172--175.

\bibitem{FBJ73}
M.~E. Fisher, M.~N. Barber, D.~Jasnow, Helicity modulus, superfluidity, and
  scaling in isotropic systems, Phys. Rev. A 8~(2) (1973) 1111--1124.

\bibitem{CL95}
P.~M. Chaikin, T.~C. Lubensky, Principles of condensed matter theory, Cambridge
  University Press, Cambridge (GB), 1995.

\bibitem{GC99}
R.~Garcia, M.~H.~W. Chan, Critical fluctuation-induced thinning of $^4${He}
  films near the superfluid transition, Phys. Rev. Lett. 83 (1999) 1187.

\end{thebibliography}

\end{document}